# Measurement of Enthalpy and Entropy of the Rate-Determining Step of a Model Electrocatalyst for the Oxygen Evolution Reaction

Joaquín Morales-Santelices[a] and Marcel Risch*[a]

[a]  Mr. J. Morales-Santelices, Dr. M. Risch
    Nachwuchsgruppe Gestaltung des Sauerstoffentwicklungsmechanismus
    Helmholtz-Zentrum Berlin für Materialien und Energie GmbH
    Hahn-Meitner Platz 1, 14109, Berlin, Germany
    marcel.risch@helmholtz-berlin.de

**Abstract:** Experimentally determined thermodynamic parameters are rarely reported for electrocatalytic reactions including the oxygen evolution reaction (OER). Yet, they contain unique and valuable mechanistic insight and present a missing link to theoretical investigations. Herein, a protocol for determining thermodynamic properties of the rate determining steps (RDS) of the OER is presented. Cobalt oxide is investigated at pH 7 as a case study. Two different approaches are employed: steady state polarization (SSP) that uses chronopotentiometry at different temperatures and current values, and potentiostatic electrochemical impedance spectroscopy (PEIS) at different DC voltages and temperatures. The data is used to fit a 3D plane from which entropy and enthalpy of the RDS are obtained. The data analysis requires an appropriate filtering of the data. Hence, we discuss suitable figures of merit for establishing appropriate filtering criteria. The values obtained are 0.72 and -0.39 eV (at 298 K) for enthalpic and entropic contributions, respectively. The obtained values are reproducible for both approaches and consistent with literature. We further highlight that the RDS entropy gives clue of the number of water molecules adsorbed or protons released prior to RDS and thereby helps identifying plausible OER mechanisms.

## Introduction

To increase the amount of hydrogen production using green sources of energy, water electrolyzers should use more durable, cheap, and active electrocatalytic materials for the water splitting reaction (WSR)[1-5]. For these purposes, it is vital to understand the mechanisms of the hydrogen and oxygen evolution reactions (HER and OER) on the surface of the electrocatalysts in a thorough way. These paths explain how to decrease the high energy of activation[6], how to improve measures of the specific activity such as the turnover frequency (TOF), and/or how to maximize the faradaic efficiency. Of the two half-reactions, the OER is the more sluggish one, and therefore, the one that needs further optimization[7].

For more than a century, the Sabatier principle[8] has been used as a guideline in heterogeneous catalysis, where it is desirable to have catalysts with binding energies (and/or other types of performance descriptors) inside a certain range that ensures that the bond is neither too strong nor too weak. In electrocatalysis of the OER, this analysis is frequently done by calculating relevant thermodynamic properties and resulting theoretical overpotential[9-10] or identifying experimental correlations[11]. Tweaking the binding energies by modifying either the catalyst composition or the OER conditions, allows us to engineer the steps undergone and encourage certain pathways while discouraging others. Transition State Theory (TST)[12] relates the difference in energy between the transition state (TS) and the reactants with a measure of the overall reaction rate. From all the different elementary reactions, the limiting one is known as the rate-determining step (RDS)[13]. Often it is not completely clear, which the RDS is, especially near the top of the volcano, where the most interesting materials are found.

When a system exhibits Tafel behavior, every Tafel region with one slope value is linked to the TS of the RDS under these conditions. In previous work of our group[14], we have simulated the influence of the surface coverage on the OER Tafel slope values and discussed how this can change with different values of the microkinetic constants of most accepted OER mechanisms in alkaline media.

Continuing with the idea of the search for cheaper, more abundant, but also more efficient and durable OER catalysts, first row transition metal oxides such as Fe, Mn, Co, Ni have outstanding electrocatalytic characteristics in neutral and alkaline media and are earth-abundant[15-17]. In the case of cobalt, there is one catalyst, amorphous cobalt oxide with molecular properties (CoP$_i$, referred to as CoCat herein), discovered by Kanan and Nocera in 2008[18-20], that has been widely studied since then[21]. This is because: it is easy to synthesize (via electrodeposition), it can drive OER in neutral media, giving better safer operating conditions, is more compatible with natural waters. Additionally, it shows self-healing[20,22].

The thermodynamic parameters of the OER for oxides containing Co can be found in the literature, mostly coming from theoretical works, including a diagram of Gibbs free energy (G) at the elementary steps of each of the four electron transfers (energy of five states in total)[23-26]. Few reports discuss the transition states between the intermediates and calculate their respective energies, which would better connect to kinetic experimental work. To fully describe the RDS of an electrocatalytic reaction, the following parameters are needed: enthalpy ($\Delta^{\ddagger}H$), entropy ($\Delta^{\ddagger}S$), Gibbs free energy ($\Delta^{\ddagger}G$) and overall transfer coefficient ($\alpha_{RDS}$). These thermodynamic parameters have been previously investigated theoretically[27-29] and experimentally[30-34].

Table 1: Enthalpies of the rate-limiting steps for OER driven in Co-oxide-containing materials found in experimental studies in the literature.

| Material | Parameter | Value | Conditions | Method | Ref. |
|---|---|---|---|---|---|
| CoCat | $\Delta^\ddagger H$ | 0.74 eV | @ η = 0.410 V (@1 mA cm$^{-2}$) 20 mV s$^{-1}$ pH = 7.0 $p_{O2}$ = 0.21 atm KPi 0.1 M 20 to 70 °C | Multi T LSV | [30] |
| | | 0.74 eV | @ η = 0.410 V KPi 0.1 M pH = 7.0 | SI-SECM | [34] |
| | $\Delta^\ddagger G$ | 0.75 eV | @ η = 0.410 V KPi 0.1 M pH = 7.0 | SI-SECM | [43] |
| LDH FeCo G-FeCo G-FeCoW A-FeCoW | $\Delta^\ddagger H$ | 0.84 eV 0.62 eV 0.51 eV 0.83 eV | @ η = 0.300 V Au(111) substrate 1 M KOH 294 to 357 20 to 80 °C | Multi T CA | [32] [32] [32] [32] |
| NiCoP NiCoFeP IrO$_2$ | $\Delta^\ddagger H$ | 0.39 eV 0.28 eV 0.31 eV | @ η = 0.700 V Au(111) substrate CO$_2$-saturated 0.5 M KHCO$_3$ 18 to 35°C | Multi T LSV | [33] [33] [33] |
| NiCo$_2$O$_4$ | $\Delta^\ddagger H^0$ | 0.72 eV | @ η = 0 1 mV s$^{-1}$ 30% w/w KOH 0° to 160°C | Multi T CA | [31] |

Theoretical investigations rely mostly on density functional theory (DFT), and sometimes combine it with molecular dynamics (MD)[29] or microkinetic simulations[35,36]. In these works, a mechanism for OER on the surface of the Co-based oxide is suggested. A Gibbs free energy profile[37] is calculated at given conditions of overpotential, pH and temperature, where a certain surface coverage is expected (according to surface Pourbaix diagrams of Co oxides)[28]. From these energy profiles, the potential determining step (PDS) can be elucidated (Figure 1a)[38]. In some works, kinetic information such as the RDS enthalpy was obtained as the difference in enthalpy of the initial state (reactants) and the RDS transition state[27,28].

In experimental investigations, the value of the RDS enthalpy is determined (even though the mechanism is not known) using either Arrhenius[31,32] or Eyring-Polanyi[30] plots where the former is more frequently used but the latter is a more straightforward treatment. These plots can be built from different types of temperature dependent experiments: linear sweep or cyclic voltammetry (LSV or CV)[30], steady state polarization (SSP) techniques such as chronoamperometry (CA) or chronopotentiometry (CP)[31-33] or electrochemical impedance spectroscopy (EIS)[17]. It is important to note, that to have a reliable result, one needs to ensure that: (1) the current measured corresponds to OER rather than other reactions or to non-faradaic current[39]. (2) the kinetics is limited by charge-transfer at these range of overpotential[40] (3) the system reaches a pseudo-steady state[41,42]. For the reasons mentioned, it is strongly recommended to prefer SSP instead of voltametric techniques and leave the applied potential for long time enough to ensure that is has arrived effectively to a pseudo-steady state. Key thermodynamic parameters for selected Co oxides are compiled in Tables 1 and 2 for experimental and theoretical studies.

In this work, we provide a protocol to experimentally determine the key thermodynamic parameters $\Delta^\ddagger H^0$, $\Delta^\ddagger S$, $\Delta^\ddagger G$ and $\alpha_{RDS}$ for the OER on CoCat as a case study. CoCat is used since it is an easy-to-prepare and widely studied material. The electrolyte is 0.1 M phosphate buffer at pH 7. To put this in context we need to start succinctly describing the theory behind it and making definitions. We propose a combined synergistic approach between SSP and potentiostatic EIS (PEIS) to tackle their individual disadvantages. Finally, we will discuss it in the context of the present literature on this electrocatalyst, its accuracy and how this could be a helpful tool for determining experimentally the suitability of proposed OER mechanisms.

Table 2: Enthalpies of the rate-limiting steps for OER driven in Co-oxide-containing materials found in theoretical studies in the literature.

Legend: E - Elementary steps energies were calculated. T - Transition states energies calculated. M: Microkinetic simulation. W – Molecular mechanics. NS – not stated.

| Material | Parameter | Value | Conditions | DS | Ref. |
|---|---|---|---|---|---|
| β-CoOOH (0001) (01$\bar{1}$2) (10$\bar{1}$4) | $\eta_{theo}$ $\eta_{theo}$ $\eta_{theo}$ | 0.80 V 0.80 V 0.48 V | 298 K 1 bar pH = 0 | PDS O* → OOH* * → OH* O* → OOH* | [26] E |
| Co$_3$O$_4$ spinel (100) Clean 0.5 O ML 0.5 OH ML | $\eta_{theo}$ $\eta_{theo}$ $\eta_{theo}$ | 1.82 V 0.45 V 0.93 V | 300 K pH = 0 | PDS OOH* → * OH* → O* O* → OOH* | [23] E |
| Co$_3$O$_4$ spinel (001) | $\Delta^\ddagger H$ $T°\Delta^\ddagger S$ $\Delta^\ddagger G$ $\eta_{theo}$ | -0.591 eV -0.671 eV +0.09 eV 0.74 V | @ η = 0.74 pH = 0 300 K | PDS OH* → O* RDS *=O → OH* | [27] ET |
| β-Co(OH)$_2$ | $k^0$ $\alpha_{RDS}$ | 1678 s$^{-1}$ 0.283 | | RDS OH* → O* | |
| β-CoOOH | $k^0$ $\alpha_{RDS}$ | 1700 s$^{-1}$ 0.507 | 298 K pH = 12.9 | OH* → O* | [28] ETM |
| CoO$_2$ | $k^0$ $\alpha_{RDS}$ | 4.98 s$^{-1}$ 0.51 | | * → OH* | |
| LaCoO$_3$ (001) SrCoO$_3$ (001) | $\eta_{theo}$ $\eta_{theo}$ | 0.46 V 0.61 V | NS NS | PDS O* → OOH* OH* → O* | [25] E |
| Co$_4$O$_4$ (cubane) | $\eta_{theo}$ $\Delta^\ddagger G$ | 0.60 V 0.15 eV | pH = 7.0 | RDS 2*=O → *OO* | [29] EW |
| Na$_2$CoP$_2$O$_7$ | $\eta_{theo}$ | 0.42 V | pH = 7.0 | OH* → O* | [24] E |
| CoCat | $\eta_{theo}$ | 0.44 V | pH = 7.0 | O* → OOH* | |

**Transition State Theory**

For OER in the overpotential zone limited by charge transfer, the current density (j) with the reaction rate (r) are related to a heterogeneous kinetic constant (k) and the surface density of active sites ($\Gamma_{act}$)[38]:

$$j = \frac{i}{A} = z_e F r = z_e e \Gamma_{act} a_{H_2O} k, \qquad (1)$$

where k, according to TST, is related to the transition state free energy and a transmission coefficient $\kappa$ by the Eyring-Polanyi equation:

$$k = \frac{\kappa k_B T}{h} exp\left(\frac{-\Delta^\ddagger G}{k_B T}\right). \qquad (2)$$

Free energy surfaces are shifted towards more negative values when a positive overpotential is applied, favoring oxidation, as sketched for OER in Figure 1b according to the relation:

$$\Delta^\ddagger G = \Delta^\ddagger G^0 - \alpha_{RDS} e\eta, \qquad (3)$$

where $\alpha_{RDS}$ is the transfer coefficient of the overall OER[40], defined as[39]:

$$\alpha_{RDS} = \frac{k_B T \ln(10)}{e} \frac{d \log_{10}(j_{OER})}{d\eta}. \qquad (4)$$

From Equation (3), the enthalpic and entropic contributions (the subscript "0" represents zero overpotential) can be separated giving[44]:

$$\Delta^\ddagger G = \Delta^\ddagger H^0 - T\Delta^\ddagger S - \alpha_{RDS} e\eta \qquad (5)$$

From Equation (5) several things can be expected given one mechanism with different intermediates: (1) Each intermediate i has an associated enthalpy $\Delta^\ddagger H_i^\circ$, entropy $\Delta^\ddagger S_i^\circ$, and transfer coefficient $\alpha_i$. (2) The intermediate i involved in the RDS is the one with the highest $\Delta^\ddagger G_i$ value at certain T and η (3) Different T and η regions can lead to different RDS.
At a given T, if two η regions have a different RDS, they will have a different Tafel slope ($b_i$):

$$b_i = \frac{k_B T \ln(10)}{\alpha_i e} \qquad (6)$$

In addition to the latter, given two different mechanisms for a reaction step: the one whose RDS has the minimum energy will be the most feasible one at that T and η.
Inserting Equation (6) in the Tafel equation for an anodic process ($\eta > 0$)[40]:

$$j = j^0 exp\left(\frac{\alpha_{RDS} e\eta}{k_B T}\right), \qquad (7)$$

where $j^0$ is the exchange current density for OER:

$$j^0 = z_e e \Gamma_{act} k^0 \qquad (8)$$

$j^0$ is related to the intrinsic rate constant and the RDS Gibbs free energy analogically to Equations (1) and (2) but evaluated at zero overpotential (Equation 3). Giving rise to the intrinsic kinetic constant $k^0$:

$$k^0 = \frac{\kappa k_B T}{h} exp\left(\frac{-\Delta^\ddagger G^0}{k_B T}\right) \qquad (9)$$

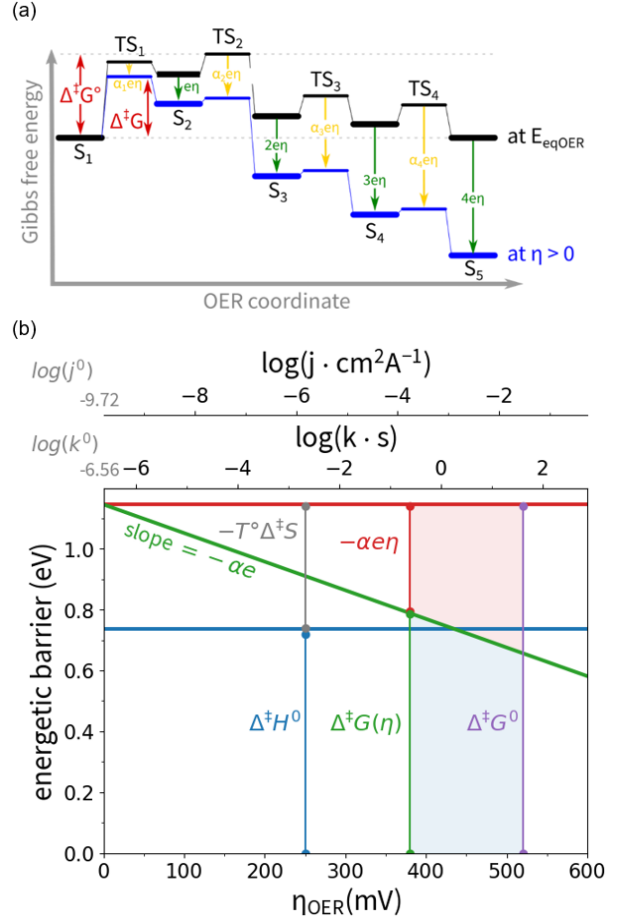

**Figure 1.** Conceptual background. (a) Gibbs free energy profile of the OER happening on an electrocatalyst's surface. The energy levels of the different states are shifted when changes in overpotential or temperature occur. $TS_i$ is rate-determining if it has the higher difference in energy relative to the reactants ($S_1$). The overall transfer coefficient $\alpha_{RDS}$ will then be equal to $\alpha_i$. (b) Interrelation between the thermodynamic parameters of the OER RDS for CoCat obtained using the values reported in the work of Lee and coworkers[30] evaluated at 25°C. The energetic barrier has different contributions (enthalpic, entropic and overpotential) and is related to the kinetic constants and current densities as stated by Equations 1 to 9. The shaded zone shows the zone where the Tafel behavior was observed. These values were extrapolated to zero overpotential to give more clarity conceptually.

## Experimental Setup

The temperature-controlled environment is a climate chamber (*Binder KB 240-UL, E6*) with a programmable temperature sequence. The chamber is at room pressure and humidity and these parameters are monitored online (*Fischer Traceable 6530*), as well as the oxygen concentration in the air (*PreSens OXY-1 SMA(-trace)-RS232-AO*). All electrochemical measurements were acquired with a Potentiostat/Galvanostat (*Gamry Instruments Reference 600+*). Electrochemical synthesis, conditioning and Tafel experiments take place on 100 mL capacity polymethyl pentene sample vials (*ALS Japan*) filled with 60 mL of solution. The electrodeposition substrate consists of fluorine-doped tin oxide (FTO) coated soda lime glass (*Ossila TEC 10, S*

*302*, 20 × 15 × 1 mm³, surface resistivity of 11-13 Ω □⁻¹) with an exposed area of 15 × 6 mm² = 0.9 cm² delimited by Kapton tape. This will be referred to as working electrode (WE). A platinum wire coil (*ALS Japan*) was used as a counter electrode (CE) both for electrodeposition and the temperature dependent routine, whereas a silver/silver chloride reference electrode (RE) immersed in saturated KCl solution (*ALS Japan RE-1BP*) was used as reference only for the latter part.

A Teflon-coated J-type thermocouple (*Sigma Aldrich*) is also inserted into the bulk of the solution to keep track of the heat transfer process between temperature steps. A reversible hydrogen electrode (RHE, *Gaskatel HydroFlex 81010*) was used to determine overpotential. A graphical description of the setup is depicted in Scheme S3.

The electrodes were cleaned by sonicating the bare FTO-coated glass in a concentrated nitric acid solution in water (13% $HNO_3$) for 10 minutes and rinsed with DI water. CE were cleaned with the same procedure.

All potentials that are reported in V vs RE refer to a saturated Ag|AgCl|KCl electrode at a mentioned temperature unless the contrary is stated. All overpotentials are reported in mV and a distinction is made between overpotential before and after $iR_u$ correction (η' and η, respectively).

CoCat film was prepared right before the temperature dependent Tafel analysis. The electrodeposition precursor (namely Electrolyte 1) was a solution of phosphate buffer 0.1 M pH = 7.0 ($KH_2PO_4$ *Sigma Aldrich*, 99.5% purity; $K_2HPO_4 \cdot 3 H_2O$ *Honeywell*, 99.5% purity) and 0.5 mM $Co^{II}$ ($Co^{II}(OH_2)_6(NO_3)_2$ *Sigma Aldrich*, 99.9% purity). The electrolytes were left inside the climate chamber to equilibrate with the atmospheric pressure at a controlled temperature of 25 °C to reach air saturation. The film synthesis consisted in a chronopotentiometry with a two-electrode configuration using a current of 12 μA cm⁻² for 750 s, when a total charge of 9.0 mC have passed (WE area is 0.90 cm², therefore the charge density is 10 mC cm⁻²).

As soon as the film was deposited, the WE with the catalyst and the CE are removed from the solution. The CE was rinsed with DI water and then dried, Electrolyte 1 rests were wiped from the WE and then both were transferred to a solution without $Co^{II}$ containing phosphate buffer 0.1 M adjusted to pH 7.0 (namely Electrolyte 2) to be conditioned using a three-electrode configuration with a RE. This process consisted of 100 voltammetric cycles starting from 0.9 V having 1.15 V and 0.40 V (all vs RE) as higher and lower limit, respectively. Except from the short lapse where the material was changed from Electrolyte 1 to Electrolyte 2, a fixed potential value was always applied on the WE to avoid degradation. The CV scan rate was 100 mV s⁻¹.

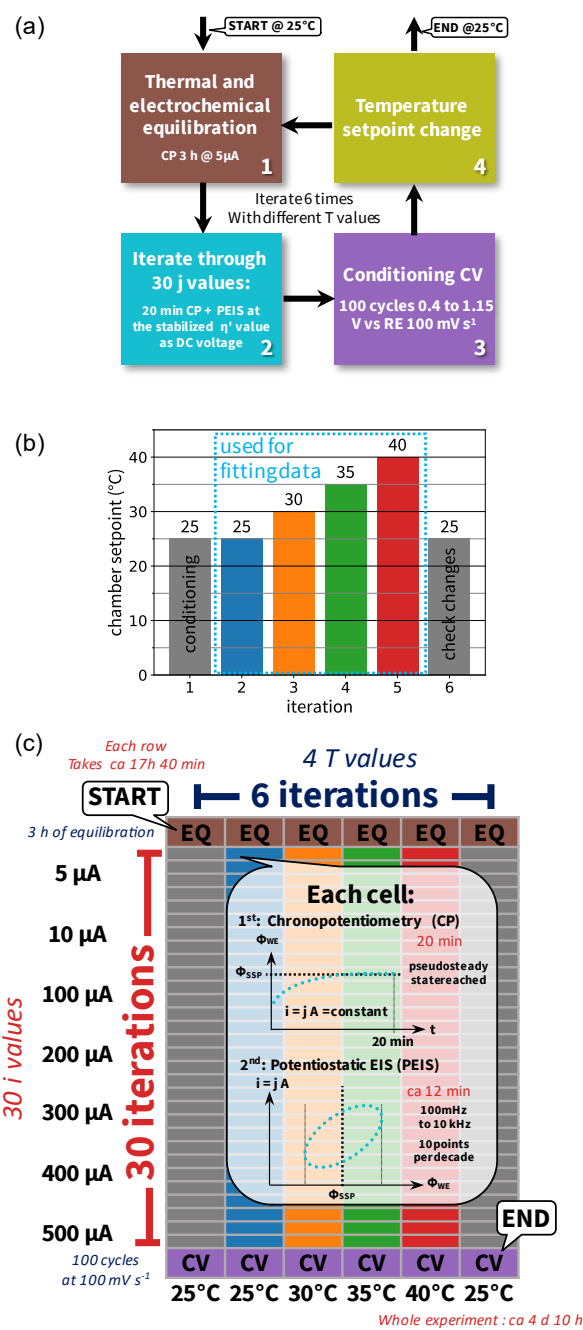

**Scheme 1.** Experimental protocol. (a) Flowchart containing the protocol used for this work. (b) Temperature steps program used in the climate chamber for these measurements. The iterative loop was repeated 6 times herein. The first and last iteration are used for conditioning and tracking irreversible changes, respectively. (c) Diagrammatic representation of the followed protocol for scanning the studied j × T region and discard the presence of irreversible changes in the material.

# Results and Discussion

In general, experimental calculation of the enthalpy of the RDS requires temperature-dependent experiments, during which the geometric and electronic structure should not change. Our protocol has an experimental part and a data treatment part as summarized in Scheme 1 and Scheme 2, respectively.

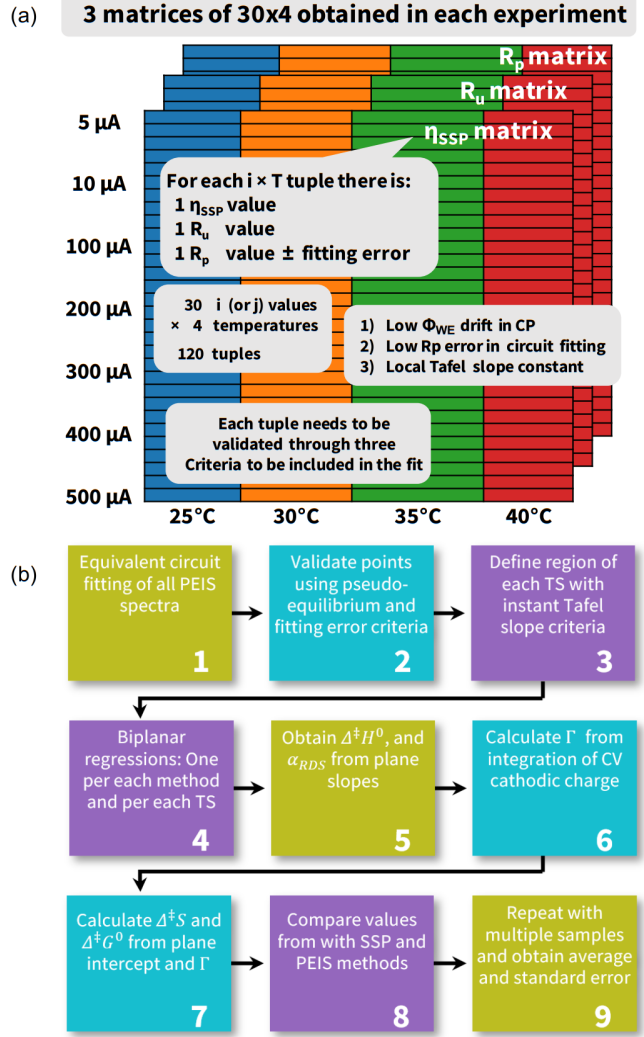

**Scheme 2.** Protocol for data analysis. (a) The experimental data obtained from the experiments explained in Scheme 1 are summarized in 3 matrices. (b) Steps followed to analyze data and obtain the thermodynamic values in this work.

## Experimental protocol

As seen in Scheme 1a, the protocol has three measurement steps: (1) equilibration (EQ), (2) CP and PEIS measurements for construction of multi-temperature Tafel plots and fitting Eyring-Polanyi equation and (3) CV to determine the redox activity. The length of the equilibrating procedures was such that could ensure that the system have reached thermal equilibrium at the beginning of each step (see details in SI Figure S1). Parameters, and raw data are accessible at [45].

After 3 h of thermal equilibration between the climate chamber and the beaker, the temperature registered with the thermocouple inside the electrolyte stabilizes and do not fluctuate during the rest of the iteration. Due to the latter, it was assumed that thermal equilibrium between the beaker and the air in the chamber was reached (Figure S2). We investigated temperatures between 25°C and 40°C in 5 °C steps. Temperatures higher than 40°C were not used because the increase in the evaporation rate of the solvent during the long experiment resulted in a sizable change in the electrolyte volume and consequently exposed area of the sample in our current setup.

## Data treatment protocol

The overpotential (without $iR_u$ compensation applied) at a given set of conditions, η', can be obtained by

$$\eta' = \phi - E \quad (10)$$

Where E is the OER equilibrium potential and ϕ is the potential applied at the WE. The relation between potential vs Ag|AgCl and overpotential at a given absolute temperature (T) in the mentioned conditions was calculated with the linear relation, which was obtained experimentally by measuring its open circuit potential (OCP) of the Ag|AgCl respect to RHE in Electrolyte 2 (Figure S1)

$$\frac{\eta'}{mV} = 1260.3 - 0.812 \frac{T}{°C} \quad (11)$$

To compensate the ohmic drop due to uncompensated resistance, in each different condition, the $R_u$ value was approximated as the real part of the impedance value with the phase angle closest to 0° in the complete range of frequencies measured (100 mHz to 100 kHz). This was achieved at higher frequencies values (above 5 Hz).

Once the full set of experiments is performed and $iR_u$ compensated, the pseudo steady-state value of current density can be obtained as the arithmetic mean of the 10 last chronoamperometric points measured (to decrease noise). Only the η x T tuples that fulfill a pseudo-equilibrium condition are used in the next step. This is to ensure that the overpotential value represent the system in pseudo-steady state. We used a maximum tolerance value permitted in the overpotential drift: within the last 10 minutes of every CP, the overpotential must not change more than ±0.65 mV. We will refer to this as Criterion 1.

To construct the Eyring-Polanyi plane, Equations 5 9,7,8 were combined, divided by T and the natural logarithm was taken. This results in a 3D plane version convenient for OER, under the assumption that $\Delta^\ddagger H°$, $\Delta^\ddagger S°$ and $\Gamma_{act}$ are invariant within temperature and current/overpotential regions (where there is one type of coverage being predominant):

$$\frac{\eta}{T} = \frac{\Delta^\ddagger H^0}{\alpha_{RDS}e}\left(\frac{1}{T}\right) + \frac{k_B}{\alpha_{RDS}e} ln\left(\frac{j}{T}\right) - \frac{\Delta^\ddagger S}{\alpha_{RDS}e} - \frac{k_B}{\alpha_{RDS}e} ln\left(z_e e\kappa \frac{k_B}{h} \Gamma_{act}\right) \quad (12a)$$

Or writing (12a) as a plane equation in 3 dimensions:

$$z = C_1 x + C_2 y + C_3 \quad (12b)$$

With variables:

$$x = \frac{1}{T} \quad (12c) \qquad y = \ln\left(\frac{j}{T}\right) \quad (12d) \qquad z = \frac{\eta}{T} \quad (12e)$$

And coefficients:

$$C_1 = \frac{\Delta^{\ddagger} H^0}{\alpha_{RDS} e} \quad (12f)$$

$$C_2 = \frac{k_B}{\alpha_{RDS} e} \quad (12g)$$

$$C_3 = -\frac{\Delta^{\ddagger} S}{\alpha_{RDS} e} - \frac{k_B}{\alpha_{RDS} e} \ln\left(z e \kappa \frac{k_B}{h} \Gamma_{act}\right) \quad (12h)$$

This will be called the SSP approach.

Analogous to what is discussed in the work of Doyle, Godwin, Brandon and Lyons[17], a partial derivative respect to η can be applied to Equation 12a. Knowing that, within the charge transfer limiting region, the faradaic resistance (per unit of area) is given by:

$$\frac{\partial j}{\partial \eta} = \frac{1}{R_f A} = \frac{1}{\tilde{R}_f} \quad (13)$$

An alternative linearized Eyring-Polanyi equation can be generated by obtaining the partial derivative respect of η in Equation (12a) and rearranging terms:

$$\ln\left(\frac{1}{\tilde{R}_f}\right) = \frac{-\Delta^{\ddagger} H^0}{k_B}\left(\frac{1}{T}\right) + \frac{\alpha_{RDS} e}{k_B}\left(\frac{\eta}{T}\right) + \frac{\Delta^{\ddagger} S}{k_B} + \ln\left(\frac{z_e \alpha_{RDS} e^2 \kappa \Gamma_{act}}{h}\right) \quad (14a)$$

3D plane equation

$$z' = C_1' x' + C_2' y' + C_3' \quad (14b)$$

with variables:

$$x' = \frac{1}{T} \quad (14c) \qquad y' = \ln\left(\frac{j}{T}\right) \quad (14d) \qquad z' = \frac{\eta}{T} \quad (14e)$$

and coefficients:

$$C_1' = \frac{-\Delta^{\ddagger} H^0}{k_B} \quad (14f)$$

$$C_2' = \frac{\alpha_{RDS} e}{k_B} \quad (14g)$$

$$C_3' = \frac{\Delta^{\ddagger} S}{k_B} + \ln\left(\frac{z \alpha e^2 \kappa \Gamma_{act}}{h}\right) \quad (14h)$$

This will be called the PEIS approach where $\tilde{R}_f$ is obtained by fitting an equivalent circuit to the EIS spectrum obtained at DC overpotential η using Gamry Analyst 2.0. Polarization resistance results obtained by equivalent circuit fitting are also filtered according to Criterion 2: Discard any data where SSP equilibrium was not fulfilled or points where the regression error calculated for the polarization resistance is greater than 50% of the $R_2$ value. Finally, the coefficients in Eq. (12) and (14) were obtained with a bilinear regression fitting of the respective experimental dataset using the Python library module *SciPy* (submodule *stats*).

To estimate the density of redox-active cobalt sites (Γ), the cathodic charge in the cyclic voltammetry of Step 4 of Scheme 1 was evaluated using the following formula:

$$\Gamma = \int_{E_0}^{E_f} \frac{1}{v z_e e} \frac{j(E)}{E} dE. \quad (15)$$

There, $v$ stands for the scan rate used. The integral boundaries ($E_0$ and $E_f$) correspond to the potential zone where the WE experienced reduction ($j(E) < 0$)[22] and $z_e$ is 2 as each Co(IV) center can be reduced until Co(II)[21].

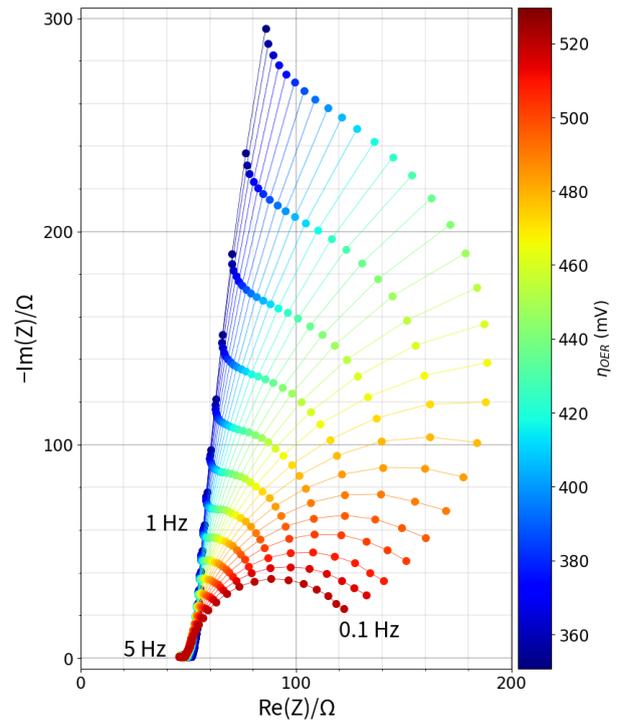

**Figure 2:** Nyquist plot of part of the PEIS spectra of CoCat electrodes at a temperature of 25°C at different overpotentials. The values chosen as DC voltage for PEIS are the pseudo-equilibrium values reached in SSP conditions in the CP. Same trend is observed for subsequent iterations at different temperatures. Continuous line connects points obtained at the same DC potential (and therefore same $\eta_{OER}$) between frequency values of 0.1 Hz (rightmost points) and 5 Hz (downmost points), 10 points per decade and 10 mV AC voltage rms. Complete PEIS (from 100 mHz to 100 kHz) is shown in Figure S3 of SI. Dataset 2349A is shown this plot. The results of fitting all the datasets to equivalent circuits can be seen in Figure S5 of SI.

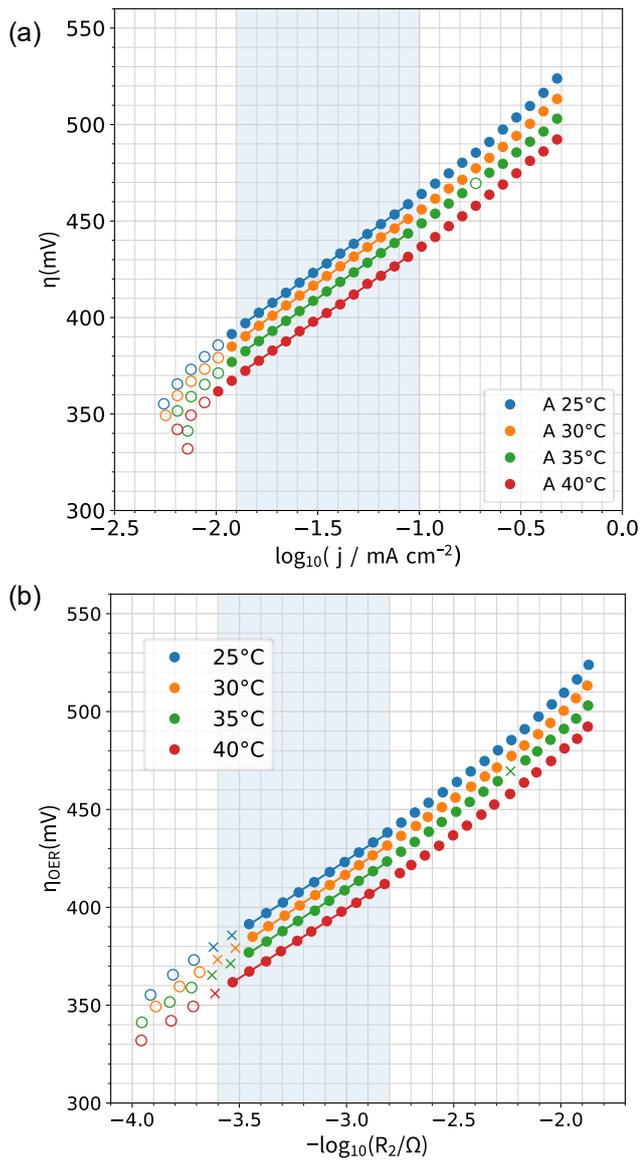

Figure 2 shows the general trend obtained at a fixed temperature with increasing DC potential at lower frequency range (100 mHz to 5 Hz). To see examples of the complete frequency range spectra refer to Figure S3 of SI.

In Figure 3, two approaches for Tafel plots are compared: (1) SSP and (2) PEIS. Even though the lines obtained from SSP appear straight by eye, the instantaneous Tafel slope (discrete derivative, analog to a previous work of our group[45]) between consecutive points increases (Figure S7). As one of the assumptions of validity is that coverage and Tafel slope are constant that region, a third criteria is necessary: the instantaneous Tafel slope on that point, that could be approximated as the forward discrete derivatives:

$$b_i \approx \frac{\eta_{i+1} - \eta_i}{\log_{10}(j_{i+1}) - \log_{10}(j_i)} = \frac{\eta_{i+1} - \eta_i}{\log_{10}(i_{i+1}) - \log_{10}(i_i)} \quad (16)$$

$$b_i \approx \frac{\eta_{i+1} - \eta_i}{\log_{10}\left(\tilde{R}_{p_i}\right) - \log_{10}\left(\tilde{R}_{p_{i+1}}\right)} = \frac{\eta_{i+1} - \eta_i}{\log_{10}\left(R_{p_i}\right) - \log_{10}\left(R_{p_{i+1}}\right)} \quad (17)$$

For SSP and PEIS approach, respectively. Criterion 3 for this dataset was defined to consider a zone to have Tafel behavior where $b_i$ is within a range of ±5 mV, namely 70 to 80 mV dec$^{-1}$, for at least 1 decade. Instantaneous Tafel plots were used to decide the span of the Tafel zone and are displayed in Figure S7. Using Criteria 1 and 3 for SSP and 1,2 and 3 for PEIS approach, the results are consistent with the expected trend of slightly increasing slope value as it is proportional to the absolute temperature according to Equation 6. The range of current density used is 10 to 100 μA cm$^{-2}$ (η values between 360 to 420 mV for SSP and 340 to 440 for PEIS approach).

**Table 3:** Ranges of Tafel slope between 25°C and 40°C, RDS enthalpy at zero overpotential and transfer coefficient obtained by SSP and EIS approaches between 25 and 40°C, and 6 and 566 mA cm$^{-2}$.

| Parameter | SSP | PEIS |
|---|---|---|
| η range (mV) | 370 to 430 (25°C)<br>390 to 460 (40°C) | 360 to 410 (25 °C)<br>390 to 440 (40°C) |
| j range (μA cm$^{-2}$) | 12 to 100 μA cm$^{-2}$ | 10 to 100 μA cm$^{-2}$ |
| b range (mV dec$^{-1}$) | 70 to 80 | 70 to 80 |
| $\alpha_{RDS}$ | 0.79 ± 0.02 | 0.85 ± 0.06 |
| $\Delta^\ddagger H^0$ (eV) | 0.72 ± 0.05 | 0.81 ± 0.06 |
| $\Delta^\ddagger S$ (meV K$^{-1}$) | -1.62 ± 0.14 | -1.32 ± 0.19 |
| $T°\Delta^\ddagger S$ (eV)* | -0.48 ± 0.04 | -0.39 ± 0.06 |
| $\Delta^\ddagger G^{0°}$ (eV)* | 1.21 ± 0.02 | 1.20 ± 0.03 |
| $\Gamma$ (cm$^{-2}$) | 1.5 · 10$^{15}$ | 1.5 · 10$^{15}$ |
| $k^0$ (s$^{-1}$)* | (3.07 ± 2.02) · 10$^{-8}$ | (5.6 ± 5.9) · 10$^{-8}$ |
| $j^0$ (A cm$^{-2}$)* | (9.84 ± 6.46) · 10$^{-11}$ | (3.92 ± 4.18) · 10$^{-11}$ |

** $T°\Delta^\ddagger S$, $\Delta^\ddagger G^{0°}$, $k^0$, $j^0$ values were calculated at $T° = 25°C = 298$ K.

**Figure 3:** Tafel plots obtained at different temperatures using the results given by: (a) SSP and (b) EIS approaches. Nomenclature: ● fulfills both Criterion 1 and 2 (potential drift < 1.10 mV and fitting error smaller than 50% of the value, respectively). ○: does not fulfill Criteria 1 (for SSP, for PEIS does not fulfill Criterion 1 or 2). ×: does not fulfill Criterion 1 but fulfills Criterion 2. +: fulfills Criterion 1 but does not fulfill Criterion 2. Points fulfilling Criterion 3 (instant Tafel slope in the range 70 to 80 mV dec$^{-1}$) have been grouped in the blue-shaded area, which indicate that they have Tafel behavior and will be used for subsequent analysis steps. Points outside the blue area do not fulfill Criteria 3 so are also excluded from the next stage of the analysis (construction of Eyring-Polanyi bilinear plots). Only dataset 2349A is shown this plot. The results of all datasets are shown in Figure S8 and S9 of SI.

## Results

The experimental procedure was repeated three times in two identical electrochemical environments that were running simultaneously inside the climate chamber, generating in total 6 independent sets of data.

**Table 4:** Criteria employed for data filtering in this work and the values used to obtain parameters shown in Table 3.

| Criterion | Description | Value | Detail |
|---|---|---|---|
| 0 | Temperature drift in EQ | < 0.05 °C | for the last 60 min |
| 1 | Overpotential drift in CP | < 0.65 mV | for the last 10 min |
| 2 | $\tilde{R}_f$ error in circuit fitting in PEIS | < 25% | of fitting value |
| 3 | Instant Tafel slope from CP or PEIS | 70-80 mV dec$^{-1}$ | this range |

As each tuple starts with 20 min of CP at a current values obtained with potentials where Co(II) experiences oxidation, most part of the sites are already oxidized when the PEIS starts. That is the reason why the contribution of this faradic process to the total charge transfer resistance is much lower when compared to the OER one using 10 mV rms as AC voltage.

The construction of an Eyring-Polanyi plot in 3D can be used to visually assess the degree of bilinearity of the data, corroborated with the determination of the goodness of fit. From the regression coefficients (either $C_1$, $C_2$, $C_3$ or $C_1'$, $C_2'$, $C_3'$) and the aid of the estimation of the film's reductive charge of $1.5 \cdot 10^{15} \ cm^{-2}$ by Equation 15, two independent set of parameters were determined (one per each approach per each transition state). These parameters were generated separately for each repetition of the experiments and statistics were obtained. The result found by the two approaches and different repetitions are presented in Table 3. The Criteria used for these values are also summarized in Figure 4. To achieve better comprehension of the combined effect of the enthalpy and the transfer coefficient on determining which TS is limiting the kinetics, a plot of Gibbs free energy versus overpotential such as the one shown in Figure 1b should be traced, as the reaction will proceed by the TS with lower energy at a given η.

**Discussion**

*Validity of equations, approximations and precautions*
The average results of the analysis were collected in Table 3 where the determined parameters were comparable among the two used methods. From all the values obtained, the transfer coefficient α$_{RDS}$ is the simplest to obtain as, in principle, it would require only one temperature (one row in Scheme 2c and Scheme 3a). As it is proportional to temperature as shown in Equation 6, the accuracy of its value is mostly limited by an accurate calibration of the thermocouple instrument (as it is related to the absolute temperature, not only to a temperature difference).

Likewise, the value of $\Delta^{\ddagger}H^0$ can be simply obtained by using only one overpotential (one column in Scheme 2c and Scheme 3a) and its accuracy mostly depends on the proper calibration of the voltage difference that has been selected between the RE and the RHE scale. In other words, α$_{RDS}$ is most sensitive to an improper calibration of the thermocouple and $\Delta^{\ddagger}H^0$ most sensitive to one of the RE in their respective absolute scales.

Regarding the entropy, $\Delta^{\ddagger}S$, it is the constant of the three that is overall most sensible to other values, as it depends both on α$_{RDS}$ and a proper estimation of Γ.

The goodness of fit of the data chosen using a bilinear fit for the Equations 12 and 17 also supports the idea that the approximation of treating parameters such as Γ and $\Delta^{\ddagger}S$ as constants within the span used to build the grid.

For an accurate overpotential calculation, a trustworthy reference must be used, and this requires considering changes with temperature of (1) the equivalence between different reference electrodes and of (2) the OER equilibrium potential itself. For (1) it is recommended to use the approach of determining the equivalence curves between the RE and a commercial RHE electrode submerged in the same solution. On the other hand, we discourage the use RHE directly as RE in the direct measurement due to our experimental evidence of showing a sluggish equilibrium. The thermal temperature coefficient ( $dE/dT$ ) obtained experimentally with our calibration (+0.235 mV/°C from Figure S1) using the Ag|AgCl|KCl RE is similar to what has been reported in the literature (+0.235 mV/°C for 1 M KCl[46]). In the case of E$_{OER}$ (in V vs RHE) vs T there is a slight difference (0.9 mV/°C[37] versus 0.81 mV/°C in our work). In our work, we observed a strong dependency of the enthalpic value with respect to the calibration chosen. As a third way of checking if our calibration was correct, the difference between NHE and saturated Ag|AgCl|KCl at 25°C is similar to what it is expected and reported (197 mV[47,48]). We conclude that our determination of the overpotentials was accurate.

For the calculations of the equilibrium potential, it is desirable to have a good control of the temperature and recording the barometric pressure in the case of air-saturated solutions by use of Henry constants that were found in the literature for the solute and its concentration[49]. It is also recommended, especially for long experiments and/or in confined places, to keep track if there is no change in the concentration of oxygen due to product accumulation, which we did not detect.

To discard the idea of any significative leakage of ions coming from the solution into the RE through the junction, the OCP value between RE and an external RHE in the same electrolyte and temperature was repeated before and after the temperature steps at a temperature of 25°C. The difference between the stabilized OCP values before and after the experiment was around 5 mV. Summing up, the expression used for conversion of V vs RE into η at different temperatures is in agreement with the literature and the conditions chosen enabled to have a good control of this parameter, demonstrated with a neglectable drift in the RE to RHE OCP value.

It is also important to check to which extent the working electrode suffers degradation during the measurement. As the experiment duration is long and involves cycling along different potentials it comprises, to some extent, changes in the structure[22]. For the sake of the purpose of this work, these changes were minimized by experimental design. The inclusion of a CV with 100 cycles is necessary at the end of each iteration as a simple and concurrent way to track irreversible changes in the material.

To assess catalytic activity and possible degradation at the end of each row of 30 CPs, the last cycle of each conditioning CV was analysed in two forms: (1) change in reductive (cathodic) charge through iterations using Equation 15 and (2) maximum current value obtained. The changes are negligible (1.0 mA cm$^{-2}$ maximum current, density of active sites from $3.78 \cdot 10^{15}$ cm$^{-2}$ to $4.19 \cdot 10^{15}$ cm$^{-2}$) as seen in Figure S6 in both curves at 25°C.

One of the most challenging parts of the data analysis was due to the high sensitivity of the results with respect to the region chosen to do the data-fitting and the filtering of points. The underlying reason it is that, for the validity of the modified presented Eyring-Polanyi equations, the selected group of points must be related to only one TS that is determining the kinetics. η x T regions that exhibit Tafel behavior are perfect for this purpose. Unfortunately, not all the OER catalysts exhibit zones with Tafel behavior, and even those who exhibit one (or more) have transition regions where two TS have non-neglectable influence on the rate of electron transfer.

Works presenting full kinetic models for heterogeneous OER are good in demonstrating that with certain combinations of microkinetic constant values, the steps possibly preceding the RDS must be in equilibrium, an assumption, which is not always fulfilled[14,44,50-54] and therefore, the analysis suggested in this work is not suitable for these conditions. Nevertheless, we would like to be emphatic with the following facts:

Not all the points obtained by SSP/PEIS approach are usable. The points should: (1) have reached the pseudo steady state by keeping the current density constant for sufficiently long time and reaching a constant (over)potential value, (2) in the case of PEIS approach, have a neglectable fitting error, (3) have neglectable diffusional limitations (4) exhibit Tafel behavior for a sufficiently large range of current (or polarization resistance) (5) be iR$_u$ compensated, (6) not be in a transition between two TS being rate-determining, since then the density of active sites, Γ, is not constant, and (7) be in thermal equilibrium.

Fact (1) is the reason why voltametric techniques are strongly discouraged for these objectives. It could be seen that, for CoCat, there are η x T tuples in which, even after 20 min of CP, η continue drifting. This could be avoided by prolonging the CP. Nevertheless, there needs to be a tradeoff between the grid density and the CP time span, as very long programs imply longer degradation of the material, and higher amount of solvent evaporated. Another more convenient option is the formulation of figures of merit that quantify how close from equilibrium the η value after 20 min is such as Criterion 1 (potential drift).

Fact (2) can be assessed by the results of the equivalent circuit fitting performed. As PEIS spectra obtained at low η values have high error for polarization resistance values, the effect that a considerable change in R$_p$ does not significantly affect the goodness of fit ($\chi^2$) value minimized.

Fact (3): The absence of mass transport limitations was ensured by: (i) using a low charge density of electrodeposition so the confinement effect in the catalyst pores is neglectable and therefore behaves as a conventional solid-liquid interface (ii) remaining in a Tafel zone (as this implies to be far from diffusional effects). This is appreciated in Figure S4 with the distribution of relaxation times (DRT) obtained for PEIS spectra with no other considerable peaks at time constants higher than the one related to charge transfer.

Fact (4): Despite only three (non colinear) points to define a plane equation, more included points improve the statistical significance (as long as they have approximately constant instant Tafel slope value). Moreover, a broader range of current values with same Tafel behavior ensured that the coverage is not changing considerably with η within itself, and we have a unique TS being the RDS. This is also related with Fact (6) which will be discussed below.

Fact (5): The accurate determination of the ohmic loss due to uncompensated resistance is also important to define the range of constant slope. As the ohmic potential loss is proportional to current, this could make the value of the slope be overestimated at higher currents if disregarded.

Fact (6): When there is more than one zone with different Tafel slope, there are two steps being RDS, separated by a transition zone, where the instant Tafel slope changes between one value and the second[14]. It is important to remark that in this zone of mixed control both TS are non-negligible and therefore, our approach will not lead to reliable results. In this case we recommend excluding any transition region since its inclusion in the fitting will generate distorted results. Additionally, two different densities of active sites are now present, with all the associated complications of determining them individually.

Fact (7): A Teflon-coated thermocouple placed in contact with the bulk of the electrolyte allows to track online the transient evolution of the heat transfer from the climate chamber into the beaker. Once the value read by the TC stabilizes one can be sure that the system has reached thermal equilibrium.

Summing up, these 7 aspects need to be taken care of, otherwise there will be a significant distortion and variance on the thermodynamic values. A simple evaluation on the quality of the data points can be done by plotting the scatter data in 3D space using auxiliary variables: x, y and z according to Equation 12 for SSP and Equation 14 for PEIS approaches and see if all the points are coplanar. In the case of having two remarkable TS's in the chosen grid the data will look like two intersecting half-planes.

*PEIS approach*
From the PEIS analysis, two processes with different time constants were noted (Figure S3): (1) at high frequencies (between 1 and 10 kHz), corresponding to the part with Re(Z) < 50 Ω in Nyquist plot of Figure 2 and (2) at low frequencies (< 10 Hz) corresponding to the greater semi circles at Re(Z) > 50 Ω. This idea is also supported by DRT plots available in Figure S4.

To keep the simplicity during the EIS fitting process, only the frequencies lower than 5 Hz were used to fit a Randles cell with constant phase element (CPE) (see details in Scheme S2). The absence of relaxation times higher than the one associated to charge transfer, supports the hypothesis of neglectable diffusional effects and a kinetically controlled electrochemical process. This fact is needed for the validity of Equations 1 and 16. Nevertheless, the values used for the determination of the compensated resistance are values obtained at high frequency (100 kHz) where the phase angles were close to 0. As expected from a faradic process, the polarization resistance decreases with increase of $\eta_{OER}$, correlated to the curvature of the lines connecting points.

PEIS measurements in CoCat at DC potentials different than OCP have been reported in the literature[53]. Works such as the one of Neerugatti, Adhikari and coworkers[53] suggest an equivalent circuit that is comprised of two time constants associated with OER and other to the contact or ionic resistance. The work of Doyle and coworkers[17] shows another circuit with three time constants that fits most of electrodes covered with passive oxides or hydrous oxides. The work of Zaharieva and coworkers[54] also provides a suitable circuit of two time constants to model the phenomenology. The mentioned equivalent circuits are shown in Scheme S3. Orazem and Tribollet[55] discuss about the mathematical equivalence between the frequency response of different physical models in circuits displayed in Scheme S3b and Scheme S3c (when diffusion effects are neglectable) and Scheme S3d (when inner oxide layer is neglectable).

Fitting more complicated models was attempted but did not give better results. As modelling the complete phenomenology of OER on CoCat is out of the scope of this work, and the only objective is to obtain a polarization resistance value as accurate as possible, it was preferred to make a cut in the frequency and only fit the points obtained at frequencies below 5 Hz and to study only one time constant, as it could be demonstrated that both do not overlap in the obtained spectra. Hence, the behavior of this portion of the spectra can be correctly represented by a much simpler Randles cell with CPE instead of capacitor achieving a much lower goodness of fit value with 4 degrees of freedom rather than the 6 required in Scheme S3b[53] (neglecting diffusion/Warburg element), or 7 in Scheme S3c[54] and Scheme S3d[17] respectively, when all the phenomena are included. This way also the convergence of the numerical system into a viable solution is considerably simpler, even in absence of a plausible initial solution guess.

The observed trends (and errors) of the resistances obtained from PEIS can be seen for each fitting in Figure S5. $R_u$ decreases with increasing overpotential, and temperature. The conditioning at the end of each iteration is responsible of a slight increase on the polarization resistance, $R_p$, at the first measurements of each iteration. This is consistent with works that describe the capacity of this material to self-heal at potentials lower than the onset of the OER[56]. $R_p$, has the expected values and changes with increasing OER overpotential with an exponential decay. As it was already mentioned, the charge is predominantly due to OER and not to oxidative changes in Co atoms without producing oxygen as the PEIS was preceded by a CP at the same DC potential. In conclusion, DRT spectra, the low chi square values obtained in the fitting and the highly similitude between the results acquired by both approaches supported that the PEIS approach can provide thermodynamic values for the RDS as well as SSP.

*Comparison of used and obtained parameters with literature*

In the work of Sprague-Klein et al[57] a similar experimental Tafel Analysis of CoCat at different temperatures between 20 and 60°C have been reported but with different research purpose. In the works of Ahn and Bard found in the literature[34,43], they calculate the kinetic constants between $Co^{III}$ and $Co^{IV}$ sites and $H_2O$: 1.2[43] and 3.2 s$^{-1}$[34] at η = 410 mV, respectively. This coverage value was obtained by surface interrogation scanning electrochemical microscopy (SI-SECM). Also, Lee and collaborators[30] have performed very similar calculations but with a LSV at 20 mV s$^{-1}$ instead of SSP.

Now we proceed to compare the values reported in the forementioned works with this one in each parameter investigated:

(1) Current density (j) range: The range used in this work spans from $10^{-4.25}$ to $10^{-2.25}$ A cm$^{-2}$ (5.5 to 555 µA cm$^{-2}$) chronopotentiometrically. 0 to 100 µA cm$^{-2}$ used for fitting. Lee and collaborators[30] study $10^{-3.5}$ to $10^{-2}$ A cm$^{-2}$ (316 µA cm$^{-2}$ to 10 mA cm$^{-2}$) and Sprague-Klein et al[57] $10^{-3.25}$ to $10^{-3}$ A cm$^{-2}$ (562 µA cm$^{-2}$ to 1 mA cm$^{-2}$), both chronoamperometrically.

(2) Temperature (T) range: This work: 25 to 40°C. Sprague-Klein et al[57] 20 and 60°C, Lee and collaborators[30] 20 to 70°C The other groups aim to higher temperatures, but the duration of their experiments is also shorter compared to ours.

(3) Overpotential range (η): This work: 360 to 440 mV at 25°C and (390 to 410 mV at 25°C for fitting). Lee and collaborators[30] 420 to 510 mV. Sprague-Klein et al[57] 270 360 mV.

(4) Tafel slope (b): This work: 68 to 78 mV dec$^{-1}$ SSP. 64 to 72 mV dec$^{-1}$ PEIS. Lee and collaborators[30] near 70 mV dec$^{-1}$. Sprague-Klein et al[57] 61 and 76 mV dec$^{-1}$. We emphasize the dependence on the current/overpotential window chosen to have a slope value higher or lower. This has also relation with the range of overpotentials chosen in [57] as its boundaries are lower than the boundaries we used or also in [30].

(5) Exchange current density ($j^0$): This work: 0.09 to 0.13 nA cm$^{-2}$ from SSP. 0.04 to 0.08 nA cm$^{-2}$ from PEIS. Sprague-Klein et al[57] 20 and 120 nA cm$^{-2}$. Lee and collaborators[30] $j^0$ = 0.19 nA cm$^{-2}$. Regarding to the x-axis intercepts of the Tafel plots ($j^0$) the value depends strongly on the way the conversion to overpotential has been done.

(6) Transfer coefficient ($\alpha_{RDS}$): This work: 0.81 to 0.84. Lee and collaborators[30] 0.94. Sprague-Klein et al[57] 0.87 to 0.95. Our $\alpha_{RDS}$ is smaller than previously reported. First, $\alpha_{RDS}$ = 0.94 is related to a Tafel slope of ca 63 mV dec$^{-1}$ at 298 K (Equation 6) whereas our instant Tafel slope range is between 70 and

- (7) Density of active sites (Γ): This work: 50 Co atoms per nm$^2$ or $5.0 \cdot 10^{15}\ cm^{-2}$. Ahn and Bard[34] 11 Co atoms per nm$^2$ or $1.1 \cdot 10^{15}\ cm^{-2}$. Our density of active sites is likely overestimated as not all Co$^{IV}$ may reduce to Co$^{II}$. Yet, the order of magnitude agrees with the previous study.
- (8) Kinetic constant extrapolated to zero overpotential ($k^0$) at 25°C: This work: 3.07· 10$^{-8}$ and 5.6 · 10$^{-8}$ s$^{-1}$ for SSP and PEIS respectively. Lee and collaborators[30] $2.7 \cdot 10^{-7}\ s^{-1}$ at η = 0. Ahn and Bard 3.2 s$^{-1}$[34A] at η = 410 mV (TOF).
- (9) Gibbs Free energy extrapolated to zero overpotential at 25°C ($\Delta^{\ddagger} G^0$): This work: 1.21 eV SSP, 1.20 eV PEIS. 1.15 eV. Ahn and Bard[34]
- (10) Enthalpy extrapolated to zero overpotential ($\Delta^{\ddagger} H^0$): This work: 0.72 eV SSP, 0.81 eV PEIS. Lee and collaborators[30] 0.74 eV.
- (11) Entropy at 25°C ($T°\Delta^{\ddagger} S^0$): This work: -0.48 eV SSP, -0.39 PEIS. Lee and collaborators[30] -0.41 eV (using Gibbs Free energy from Ahn and Bard).

80 mV dec$^{-1}$ at temperatures between 25 and 40°C due to Criterion 3 and for this range using the same equation and temperature $\alpha_{RDS}$ = 0.81 (73 mV dec$^{-1}$) and 0.84 (70 mV dec$^{-1}$) for SSP and PEIS respectively.

Another aspect worthy mentioning is that the effect that the entropic contribution has to the energetic barrier of OER is considerable compared to the total Gibbs free energy part in both sets (35% using Lee and coworkers' values, 39.6% and 32.5% using our two approaches). In our case we used 50 Co atoms per nm$^2$ as a coverage value. In comparison with SI-SECM, we would like to remark the fact of the simplicity of the CV as a method to estimate the reductive charge for transition metal based electrocatalysts where SI-SECM analysis is not available, and the attenuated effect this would have in the precision of the entropy obtained, as in Equations 12h and 14h the coverage is inside a logarithm. In summary, our determined values agree mostly with literature and reasons for deviations have been discussed.

To finalize the discussion, we would like to remark that the determination of the entropy of the RDS is also an insightful tool, as it provides additional information to determine plausible mechanisms for OER at different conditions: as water molecules, protons and hydroxide ions have much higher entropy values compared to species adsorbed to the surface (due to the latter have less degrees of freedom, discussion can be found here[58]). The experimental value tells how many water molecules have been adsorbed or how many protons had been released prior to RDS with the help of the computational hydrogen electrode approach (CHE)[9,59]. We postulate the following relation:

$$T°\Delta^{\ddagger} S^0 \approx -0.70\ eV \cdot w + 0.20\ eV \cdot p \quad (18)$$

Where w is the number of water molecules adsorbed and p the number of protons released prior the RDS. The value -0.70 eV corresponds to the entropic contribution to Gibbs free energy at 298 K according to theoretical works such as [27]. The value of 0.20 eV is one half of the entropic contribution of a hydrogen molecule, which is equivalent to the one of one electron and one proton, also at 298 K, using the CHE.

Applying this relation to our catalyst, -0.48 and -0.39 eV, the experimentally obtained values, are consistent with the entropy change involved in the adsorption of a water molecule (-0.70 eV) plus the release of a proton-electron pair (one half of the entropy of +0.4 eV): $-0.48\ eV \approx -0.70\ eV + 0.5 \cdot 0.4\ eV$ analogously to the value that Lee and coworkers have published, using their reported values for water and hydrogen[30].

## Conclusion

In this work we present two ways of determining the thermodynamic values of the RDS for OER in a model catalyst: The SSP approach and the PEIS approach. The first one uses series of chronopotentiometries at different current and temperature values. The second one uses potentiostatic EIS measurements at different DC potentials and different temperatures. The datasets from both approaches were fitted by modified a tridimensional Eyring-Polanyi equation using biplanar regression. From the slopes of the fits, we obtained the enthalpy of the RDS and its associated transfer coefficient. Likewise, we determined the entropy of the RDS from the intercept in these plots, knowing a priori the surface density of active sites, which was herein approximated by integrating the reductive charge from CV. We discussed the main sources of error in the data analysis procedure. The most important factor was (pseudo) steady state of the system, which implies that the temperature and overpotential (in case of using chronopotentiometry) were stable. The second most important factor was an accurate choice of a region with Tafel behavior. We proposed four criteria for filtering the experimental dataset to mitigate these sources of error.

The enthalpy and entropy values that we obtained for CoCat at pH = 7 are consistent with values obtained in the literature. We also proposed that they can be used as an additional tool to determine the number of protons released and/or water molecules adsorbed per turnover prior the RDS in OER, helping to combine theoretical works with experimental ones with the scope of finding better materials for WSR. Our protocol can be readily extended to other electrocatalysts and electrocatalytic reactions and we hope that it generates data that can be readily compared to theoretical kinetics studies.

## Supporting Information

The authors have cited additional references within the Supporting Information.[60,61]

## Data Availability Statement

The data that support the findings of this study are openly available in Zenodo at https://doi.org/10.5281/zenodo.10222127, reference number 10222127.

## Acknowledgements

This project has received funding from the European Research Council (ERC) under the European Union's Horizon 2020 research and innovation program under grant agreement No 804092.

The authors declare no conflict of interest.

**Keywords:** OER rate-limiting steps • OER rate-determining step enthalpy • OER mechanism • Eyring-Polanyi • cobalt oxides.

# Appendix

Table A1. Symbols used in this work and their meaning.

| Symbol | Meaning | Value | Unit |
|---|---|---|---|
| $A$ | geometric area of the WE | | $cm^2$ |
| $a_{H_2O}$ | activity of water | | − |
| $b_i$ | Tafel slope related to $TS_i$ | | $mV\ dec^{-1}$ |
| $C_1$ | bilinear fitting constant for galvanostatic SSP (x slope) | | $V$ |
| $C_1'$ | bilinear fitting constant used for PEIS (x' slope) | | $K$ |
| $C_2$ | bilinear fitting constant for galvanostatic SSP (y slope) | | $V\ K^{-1}$ |
| $C_2'$ | bilinear fitting constant used for PEIS (y' slope) | | $K\ V^{-1}$ |
| $C_3$ | bilinear fitting constant for galvanostatic SSP (z intercept) | | $V\ K^{-1}$ |
| $C_3'$ | bilinear fitting constant used for PEIS (z' axis intercept) | | − |
| $E$ | equilibrium potential | | $V\ vs\ RE$ |
| $e$ | elementary charge unit | $1.60 \cdot 10^{-19}$ | $C$ |
| $F$ | faraday constant | $96\ 485$ | $C\ mol^{-1}$ |
| $G$ | Gibbs free energy | | $eV$ |
| $h$ | Planck constant | $4.14 \cdot 10^{-15}$ | $eV\ s$ |
| $i$ | current | | $\mu A$ |
| $i_{OER}$ | faradaic current related to OER | | $\mu A$ |
| $j$ | current density | | $\mu A\ cm^{-2}$ |
| $j_{OER}$ | faradaic current density related to OER | | $\mu A\ cm^{-2}$ |
| $j^0$ | exchange current density (extrapolated to η = 0) | | $\mu A\ cm^{-2}$ |
| $k$ | heterogeneous kinetic constant | | $s^{-1}$ |
| $k^0$ | heterogeneous kinetic constant (extrapolated to η = 0) | | $s^{-1}$ |
| $k_B$ | Boltzmann constant | $8.62 \cdot 10^{-5}$ | $eV\ K^{-1}$ |
| $p$ | protons transferred before RDS | | − |
| $R_p$ | polarization resistance | | $\Omega$ |
| $\tilde{R}_p$ | polarization resistance (normalized by area unit) | | $\Omega\ cm^{-2}$ |
| $R_u$ | uncompensated resistance | | $\Omega$ |
| $r$ | heterogeneous reaction rate | | $mol\ cm^{-2}s$ |
| $T$ | temperature | | $K$ |
| $w$ | water molecules transferred before RDS | | − |
| $x$ | auxiliary variable for galvanostatic PEIS fitting | | $K^{-1}$ |
| $x'$ | auxiliary variable for galvanostatic SSP fitting | | $K^{-1}$ |
| $y$ | auxiliary variable for galvanostatic PEIS fitting | | - |
| $y'$ | auxiliary variable for galvanostatic SSP fitting | | - |
| $z$ | auxiliary variable for galvanostatic PEIS fitting | | $V\ K^{-1}$ |
| $z'$ | auxiliary variable for galvanostatic SSP fitting | | $V\ K^{-1}$ |
| $z_e$ | number of electrons transferred per OER turnover | 4 | |
| $\Gamma_{act}$ | surface density of active sites | | $cm^{-2}$ |
| $\Delta^\ddagger G^0$ | Gibbs free energy of the RDS at η = 0 | | $eV$ |
| $\Delta^\ddagger G$ | Gibbs free energy of the RDS at a given η | | $eV$ |
| $\Delta^\ddagger H^0$ | enthalpy of the RDS at zero overpotential | | $eV$ |
| $\Delta^\ddagger H^0$ | enthalpy of the RDS at a given η | | $eV$ |
| $\Delta^\ddagger S$ | entropy of the RDS (invariant with η) | | $eV\ K^{-1}$ |
| $\alpha_{RDS}$ | transfer coefficient of the RDS | | − |
| $\alpha_i$ | transfer coefficient of the i-th electron transfer | | − |
| $\phi$ | WE potential after iRu correction | | $V\ vs\ RE$ |
| $\eta$ | overpotential after iRu correction | | $mV$ |
| $\eta'$ | overpotential without iRu correction | | $mV$ |
| $\eta_{theo}$ | theoretical overpotential | | $mV$ |
| $\kappa$ | transmission coefficient (TST) | | − |
| $v$ | scan rate | | $mV\ s^{-1}$ |
| $\chi^2$ | goodness of fit | | − |

Table A2. Abbreviations used in this work and their meaning.

| Abbreviation | Meaning |
|---|---|
| AC | Altern current |
| CA | chronoamperometry |
| CE | counter electrode |
| CHE | Computational hydrogen electrode |
| CoCat | Amorphous cobalt oxide phosphate catalyst |
| CP | chronopotentiometry |
| CPE | Constant phase element |
| CV | cyclic voltammetry |
| DC | Direct current |
| DFT | Density functional theory |
| DRT | Distribution of relaxation times |
| EIS | electrochemical impedance spectroscopy |
| EQ | Equilibration step |
| FTO | Fluorine-doped tin oxide |
| HER | hydrogen evolution reaction |
| LSV | Linear sweep voltammetry |
| MD | Molecular dynamics |
| NHE | Normal hydrogen electrode |
| OCP | open circuit potential |
| OER | oxygen evolution reaction |
| PDS | potential determining step |
| PEIS | potentiostatic electrochemical impedance spectroscopy |
| RDS | rate determining step |
| RE | reference electrode |
| RHE | reversible hydrogen electrode |
| S | state |
| SI | Supporting information |
| SI-SECM | Surface interrogation scanning electrochemical microscopy |
| SSP | Steady state polarization |
| TOF | Turnover frequency |
| TS | transition state |
| TST | transition state theory |
| WE | working electrode |
| WSR | water splitting reaction |

# Measurement of Enthalpy and Entropy of the Rate-Determining Step of a Model Electrocatalyst for the Oxygen Evolution Reaction

Joaquín Morales-Santelices[a] and Marcel Risch*[a]

[a]    Mr. J. Morales-Santelices, Dr. M. Risch
       Nachwuchsgruppe Gestaltung des Sauerstoffentwicklungsmechanismus
       Helmholtz-Zentrum Berlin für Materialien und Energie GmbH
       Hahn-Meitner Platz 1, 14109, Berlin, Germany
       marcel.risch@helmholtz-berlin.de

# Supporting Information

## S1: Deduction of equations used for PEIS approach.

The classic form of the Tafel equation[40] is:

$$\eta = \log_{10}(j^0) + b \log_{10}(j) \tag{S1}$$

Written in exponential form:

$$j = j^0 \cdot \exp\left(a \cdot \frac{\eta}{b}\right), \tag{S2}$$

If we take the derivative on both sides with respect to $\eta$ using Equation 13:

$$\frac{\partial j}{\partial \eta} = \frac{1}{R_f A} = \frac{1}{\tilde{R}_f} \tag{13}$$

And rearranging terms, we arrive to the expression shown in Lyons et al[17] for an alternative Tafel plot analysis. Equation S3a was used in Figure 3b for extracting Tafel slope and exchange current density using PEIS approach: (a = ln(10))

$$\log_{10}\left(\frac{1}{\tilde{R}_f}\right) = \frac{\eta}{b} + \log_{10}\left(\frac{aj^0}{b}\right) \tag{S3a}$$

$$\ln\left(\frac{1}{\tilde{R}_f}\right) = \frac{a\eta}{b} + \ln\left(\frac{aj^0}{b}\right) \tag{S3b}$$

Substituting both $b_i$ from Equation 6 (of the main text):

$$b_i = \frac{k_B T \ln(10)}{\alpha_i e} \tag{13}$$

And $j^0$ from Equation 8 (of the main text):

$$j^0 = z_e e \Gamma_{act} k^0 \tag{14}$$

In Equation S3b, results in:

$$\ln\left(\frac{1}{\tilde{R}_f}\right) = \frac{a\eta}{k_B T a/\alpha_{RDS} e} + \ln\left(\frac{aze\Gamma_{act}k_0}{k_B T a/\alpha_{RDS} e}\right) \tag{S4}$$

Inserting $k^0$ from Equation 9 (of the main text):

$$k^0 = \frac{\kappa k_B T}{h} \exp\left(\frac{-\Delta^\ddagger G^0}{k_B T}\right) \tag{15}$$

And rearranging terms one arrives to:

$$\ln\left(\frac{1}{\tilde{R}_f}\right) = \frac{\alpha_{RDS} e \eta}{k_B T} + \ln\left(\frac{\alpha_{RDS} e z e \Gamma_{act}}{k_B T}\right) + \ln\left(\frac{\kappa k_B T}{h}\right) - \frac{\Delta^\ddagger G^0}{k_B T} \tag{S5}$$

Decomposing $\Delta^\ddagger G^0$ into its enthalpic and entropic components one arrives to the two final expressions:

$$\ln\left(\frac{1}{\tilde{R}_f}\right) = \frac{\alpha_{RDS} e \eta}{k_B T} + \ln\left(\frac{\alpha_{RDS} e^2 z \Gamma_{act} \kappa}{h}\right) - \frac{\Delta^\ddagger H^0}{k_B T} + \frac{\Delta^\ddagger S}{k_B} \tag{S6a}$$

$$\log_{10}\left(\frac{1}{\tilde{R}_f}\right) = \frac{\alpha_{RDS} e \eta}{a k_B T} + \log_{10}\left(\frac{\alpha_{RDS} e^2 z \Gamma_{act} \kappa}{h}\right) - \frac{\Delta^\ddagger H^0}{a k_B T} + \frac{\Delta^\ddagger S}{a k_B} \tag{S6b}$$

## S2: Electrochemical details

### (A) Potentiostat parameters

Potentiostats were chassis grounded during all the measurements. The following parameters were applied:

**Table S1:** Detailed electrochemical parameters.

| Measurement | Gamry Wizard Parameter | Used Value | unit |
|---|---|---|---|
| OCP (1) | Sample period | 0.5 | s |
| | Stability | 0 | mV s$^{-1}$ |
| | Sample area | 0.9 | cm$^2$ |
| ED (2) | Pre-step current | 12.5 | µA |
| | Pre-step delay time | 0 | s |
| | Step 1 current | 12.5 | µA |
| | Step 1 time | 0 | s |
| | Step 2 current | 12.5 | µA |
| | Step 2 time | 0 | s |
| | Sample period | 0.5 | s |
| | Lower limit | 0.3 | V |
| | Higher limit | 1.5 | V |
| | Decimate | off | |
| | Sample area | 0.9 | cm$^2$ |
| | Sampling mode | Noise reject | |
| | Equilibrium time | 5 | s |
| CP (3) | Pre-step current | 5 | µA |
| | Pre-step delay time | 0 | s |
| | Step 1 current | 5 | µA |
| | Step 1 time | 10800 | s |
| | Step 2 current | 5 | µA |
| | Step 2 time | 0 | s |
| | Sample period | 1 | s |
| | Lower limit | 0.3 | V |
| | Upper limit | 1.2 | V |
| | Decimate | off | |
| | Sample area | 0.9 | cm$^2$ |
| | Sampling mode | Noise reject | |
| | Equilibrating time | 5 | s |
| CV (4) | Initial voltage | 0.4 | V vs RE |
| | Scan limit 1 | 0.4 | V vs RE |
| | Scan limit 2 | 1.5 | V vs RE |
| | Final voltage | 0.7 | V vs RE |
| | Scan rate | 100 | mV s$^{-1}$ |
| | Step size | 10 | mV |
| | Cycles | 100 | - |
| | Sample area | 0.9 | cm$^2$ |
| | I/E Range mode | fixed | |
| | Sampling mode | surface | |
| | Max current | 6 | mA |
| | Equilibrating time | 5 | s |
| | IR compensation | none | |
| CP (5a) | Pre-step current | 5-500 | µA |
| | Pre-step delay time | 0 | s |
| | Step 1 current | 5-500 | µA |
| | Step 1 time | 1200 | s |
| | Step 2 current | 5-500 | µA |
| | Step 2 time | 0 | s |
| | Sample period | 1 | s |
| | Lower limit | 0.3 | V |
| | Upper limit | 1.2 | V |
| | Decimate | off | |
| | Sample area | 0.9 | cm$^2$ |
| | Sampling mode | Noise reject | |
| | Equilibrating time | 5 | s |
| EIS (5b) | DC voltage | Last value obtained in 5a | V vs RE |
| | Initial frequency | 100 000 | Hz |
| | Final frequency | 0.1 | Hz |
| | Points/decade | 10 | |
| | AC voltage | 10 | mV rms |
| | Sample area | 0.9 | cm$^2$ |
| | Estimated Z | 32 | Ω |
| | Optimized for | low noise | |
| | THD | off | |
| | Drift correction | off | |

**Table S2:** Additional potentiostat parameters used for the experiments in this work.

| Additional Parameter | Used Value |
|---|---|
| Chassis grounding | yes |
| Cell state mode | OCP after measurements |

**(B) RE details**

Ag|AgCl|KCl 3 M electrodes (*ALS Japan RE-1CP*) were used for all the measurements. The potential conversion between RHE and Ag|AgCl|KCl 3 M was determined experimentally by measuring OCP of each RE compared to a Hydrogen Reference Electrode (*Gaskatel Hydroflex 88010*) in the same buffer used for experiments (potassium phosphate 0.1 M pH = 7.0). OCP was measured for two hours. This was repeated in each of the temperature values where measurements were done. The curve and their respective linear fitting are displayed in Figure S1.

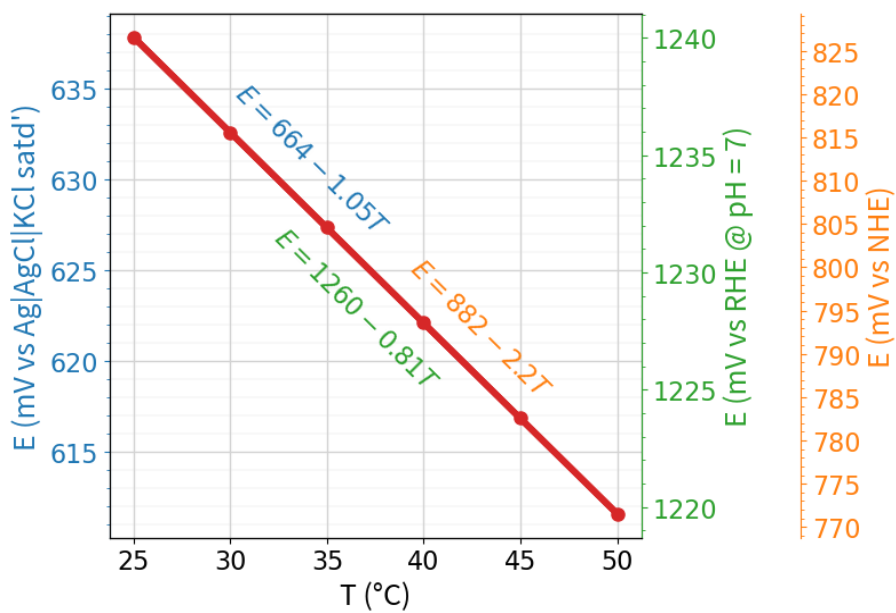

**Figure S1:** OER Equilibrium potential at different temperatures used to obtain WE overpotential in this work. This curve (blue axis) was obtained by subtraction of: (i) the experimental stabilized value of OCP between RHE and saturated Ag|AgCl|KCl electrodes used in the experiments in air saturated Electrolyte 1 at pH = 7 (not shown in the plot) and (ii) the Nernst potential for OER vs RHE at pH = 7, $pO_2$ = 0.21 atm and the respective temperature (green axis).

**Overpotential calculation at non-standard temperature, oxygen pressure and pH values**

According to the following reaction for OER:

$$4H^+_{(aq)} + 4e^- + O_{2(g)} \rightleftharpoons 2H_2O_{(l)} \tag{S7}$$

The expression for overpotential at a temperature T and an oxygen partial pressure $p_{O_2}$ and pH values is given by:

$$\eta_{OER}(T, p_{O_2}, pH) = \phi - E_{eq}(T, p_{O_2}, pH) \tag{S8}$$

Where the second term comes from the Nernst equation applied to (S7):

$$E_{eq}(T) = E_{OER}°(T) - \frac{RT\ln(10)}{4F} \cdot \log_{10}\left(\frac{a^2_{H_2O}}{a^4_{H^+} p_{O_2}}\right) \tag{S9}$$

Making pH appear explicitly:

$$E_{eq}(T) = E_{OER}°(T) - \frac{RT\ln(10)}{4F} \cdot \log_{10}\left(\frac{a^2_{H_2O}}{p_{O_2}}\right) - \frac{RT\ln(10)}{F} pH \tag{S10}$$

Converting the equilibrium potential at standard activity values $E_{OER}°(T)$ to the one at standard temperature[60]:

$$E_{OER}°(T) = E_{OER}°(T°) + \frac{dE}{dT}(T - T°) \tag{S11}$$

**Table S3:** Example of calculation of OER equilibrium (Nernst) potential at air saturated aqueous solution of pH 7.0 at 30°C.

| Parameter | Expression | Value | Unit |
|---|---|---|---|
| temperature | T | 303.15 | K |
|  |  | 30 | °C |
| oxygen pressure | $p_{O_2}$ | 0.21 | atm |
| pH | pH | 7.00 | - |
| gas constant | R | 8.31 | J mol$^{-1}$ K$^{-1}$ |
| Faraday constant | F | 96485 | C mol$^{-1}$ |
| number of electrons per OER turnover | z | 4 | - |
| standard temperature | T° | 298.15 | K |
|  |  | 25 | °C |
| standard OER potential at 25°C[61] | $E_{OER}°(T°)$ | 1.23 | V vs NHE |
| OER temperature coefficient[61] | $\dfrac{dE}{dT}$ | -0.846 | mV K$^{-1}$ |
| **Potential contributions** |  | **Value** | **Unit** |
| shift due to non-standard temperature | $\dfrac{dE}{dT}(T - T°)$ | -4.2 | mV |
| shift due to O$_2$ activity term | $-\dfrac{RT\ln(10)}{4F}\log_{10}(p_{O_2})$ | +10.2 | mV |
| **Nernst potential at T** | $\mathbf{E_{OER}(T)}$ | **1236** | **mV vs RHE** |
| shift due to pH effect term at T | $-\dfrac{RT\ln(10)}{F}\text{pH}$ | -421 | mV |
| shift from experimental curve at T | (RHE to RE) | -603 | mV |
| **Nernst potential at T** | $\mathbf{E_{OER}(T)}$ | **815** | **mV vs NHE** |
|  |  | **633** | **mV vs RE** |

## S3: Experimental Setup additional information

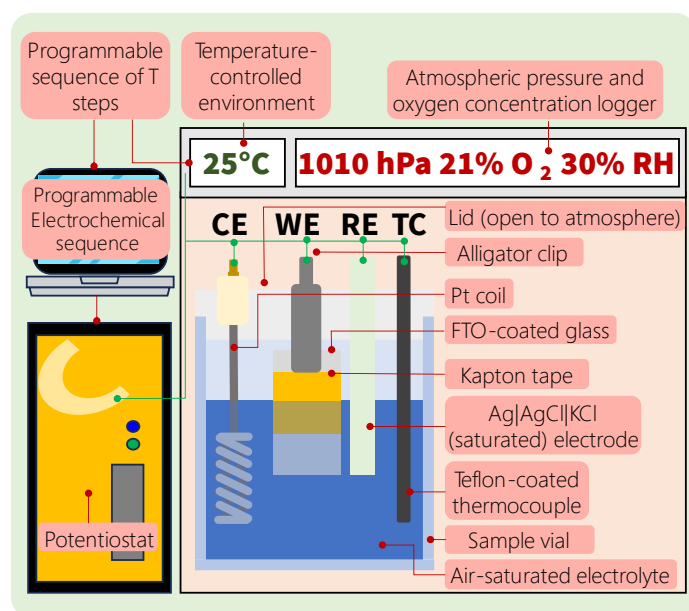

**Scheme S1:** Experimental setup: For electrodeposition a 2-electrode configuration was used. For all the subsequent experiments the represented 3-electrode configuration was used in a 100 cc sample vial with ca. 60 mL electrolyte and in contact with the atmosphere inside of the climate-chamber. The geometric exposed area of FTO in the electrolyte was 0.90 cm$^2$ and the Kapton tape limited this area, avoiding changes in the exposed area, especially at higher temperatures, due to solvent evaporation. The separation between WE and CE was near to 3 cm and from WE to RE was near to 1 cm.

## S4: Temperature monitoring during the measurements

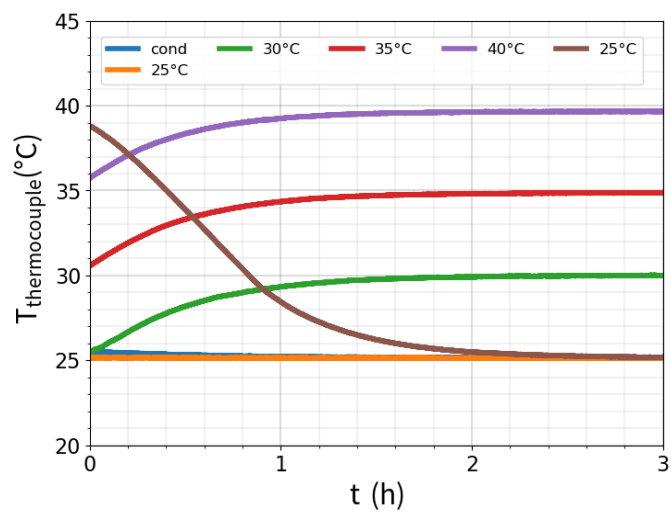

**Figure S2:** Cooling and heating curves of the 3-electrode configuration cell during Step 3 according to Scheme 1a. The curves show transient heat transfer by convection between the air climate chamber (whose setpoint changes in each iteration as shown in Scheme 1b) and the cell until thermal equilibrium. t = 0 is the moment where the setpoint changes to the value shown in the legend. Y axis shows the temperature inside the electrolyte measured by the Teflon-coated thermocouple. All original values read in the TC were shifted +0.7 C° to match climate chamber controller's temperature. During the last 12 min the temperature drift is less than 0.25 C° for all the displayed curves. Dataset 2352B was used for this plot.

**S5: PEIS approach - additional plots**

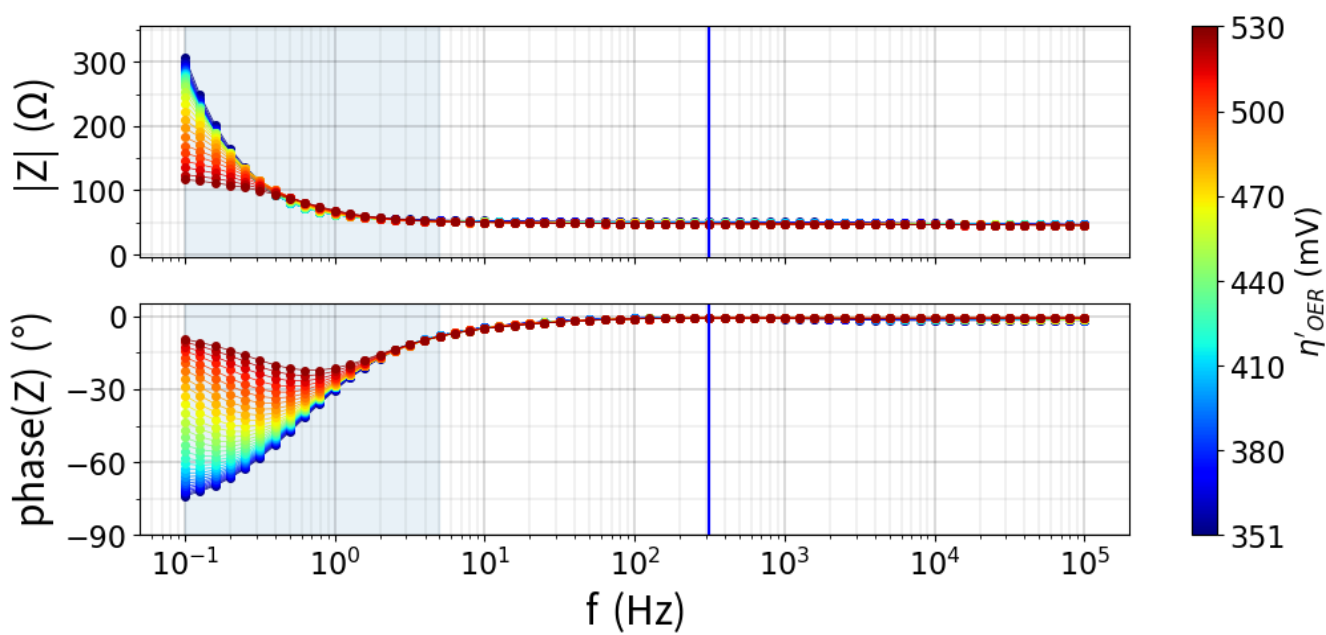

**Figure S3:** Bode plots of the complete frequency range used (100 mHz to 100 kHz) of different η' (no compensation applied) at 25°C. The shaded area shows the reduced frequency range used for the fitting of Scheme S3 (100 mHz to 5 Hz). The vertical blue line shows the frequency value where the maximum phase value (near to 0°) was obtained. The real part associated with this frequency value is $R_u$ used for $iR_u$ compensation). These plots contain dataset 2349A.

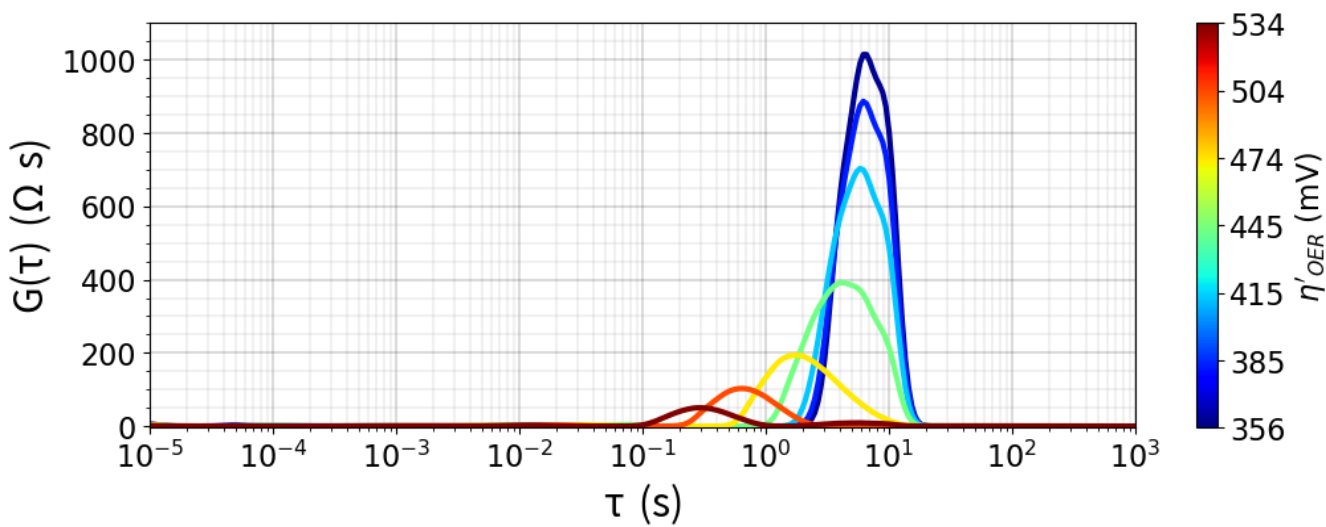

**Figure S4:** DRT fits for PEIS spectra (100 mHz to 100 kHz) obtained at 7 different η' DC overpotential values (before $iR_u$ compensation) at 25°C. In all of them the predominant peak in the order of $10^{-1}$ to $10^0$ s (blue-shaded zone) is present and associated to OER. There is an overlap of a smaller peak (whose contribution decreases with increasing η' values) in the red-shaded zone of lower frequency. Dataset 2349A was used for this plot.

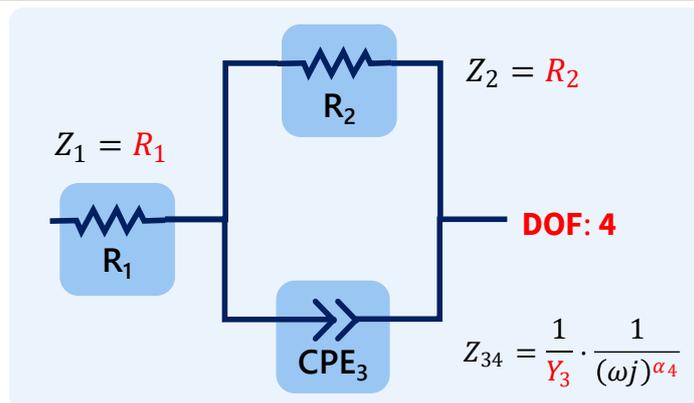

**Scheme S2:** Equivalent circuit used to fit PEIS data. $R_1$ represents the uncompensated resistance of the cell. $R_2$ represents charge-transfer resistance related to the oxidation of Co atoms in CoCat necessary to initiate OER. $CPE_3$ is the double-layer capacitance of CoCat-electrolyte interface including non-idealities. The circuit has 4 degrees of freedom (DOF), and the values fitted are enumerated from 1 to 4 and shown with their associated impedance formula ($\omega$ is angular speed in rad/s and j is the imaginary unit).

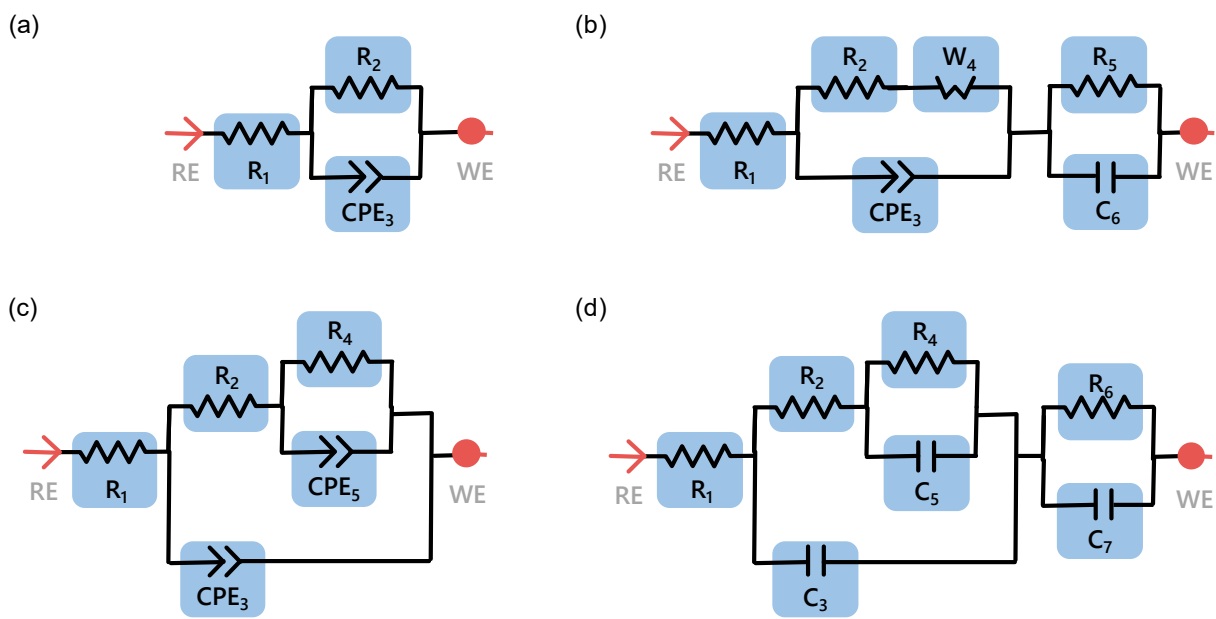

**Scheme S3:** Suitable equivalent circuits found in the literature to fit the obtained PEIS spectra. (a) The chosen one (b) Neerugatti, Adhikari and coworkers[55] (c) Zaharieva and coworkers[56] (d) Lyons group[17]. R: resistor, C: capacitor, CPE: constant phase element, W: infinite Warburg element.

S9

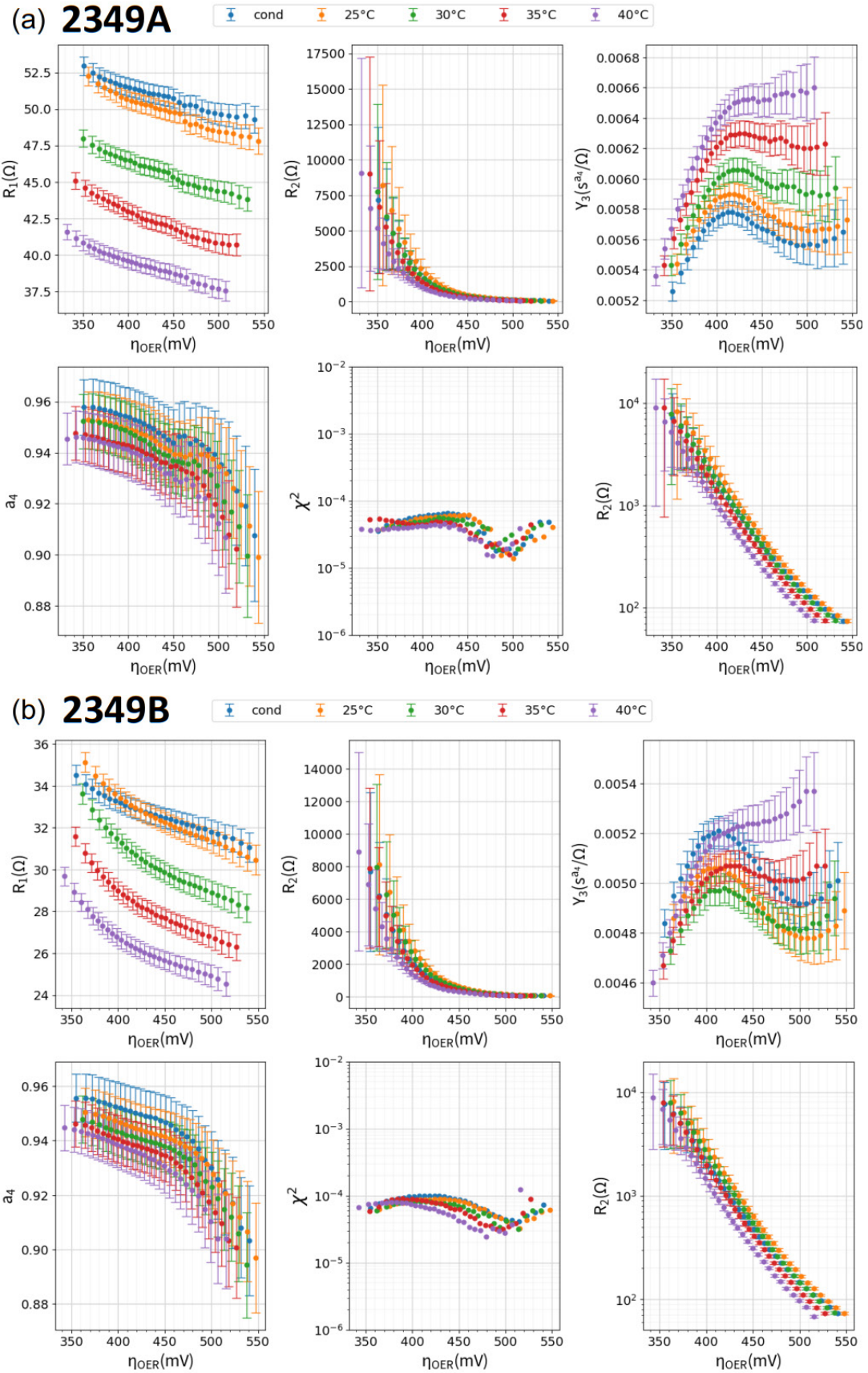

**Figure S5:** Equivalent circuit fitting result for all PEIS spectrum and goodness of fit obtained using the circuit represented in Scheme S3. Error bars represent the fitting error of each degree of freedom given by *Gamry Analyst 2* software. Points that have an error higher than 50% of the obtained value are discarded for fitting.

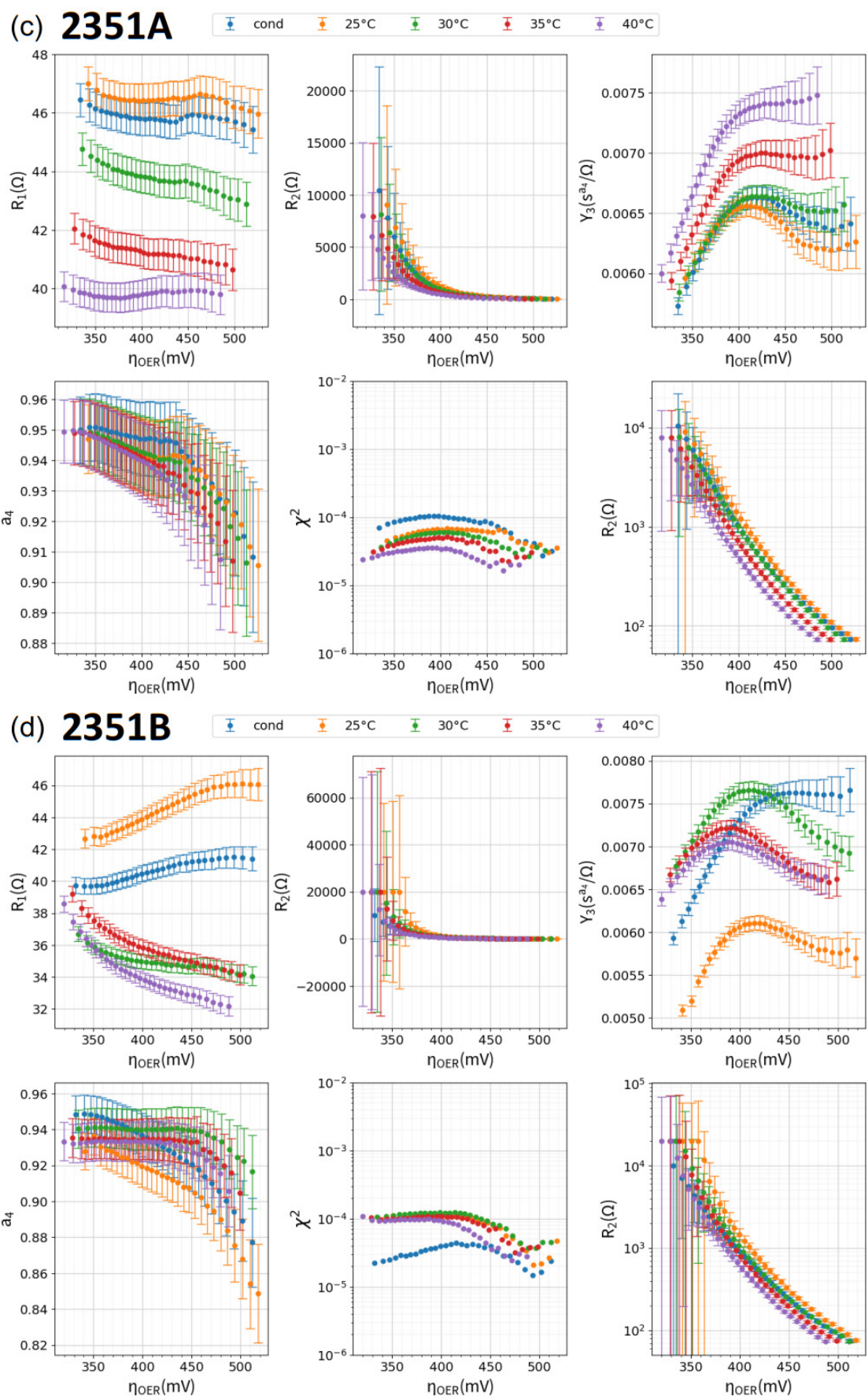

**Figure S5:** (continuation)

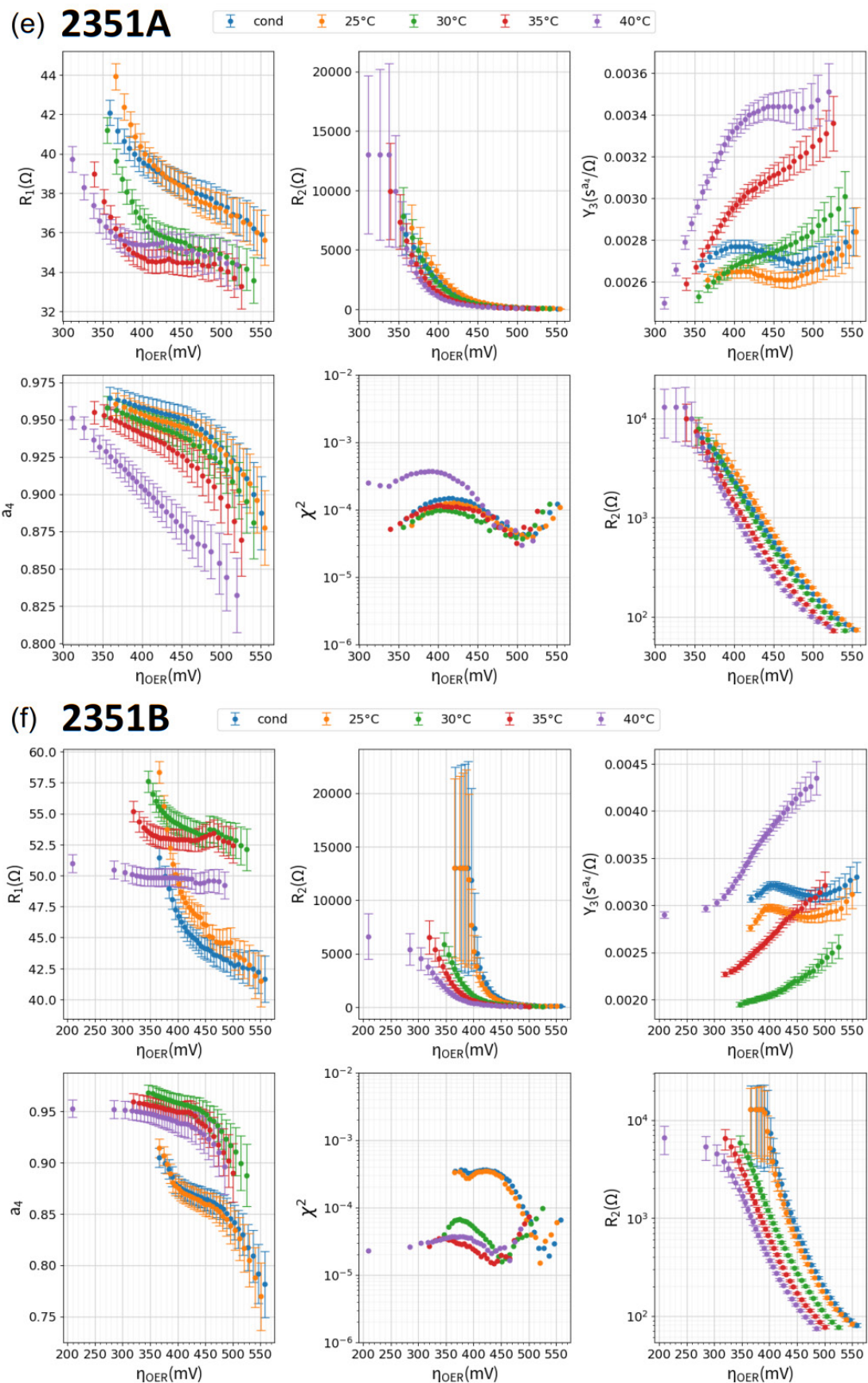

**Figure S5:** (continuation)

**S6: CV run at the end of each iteration**

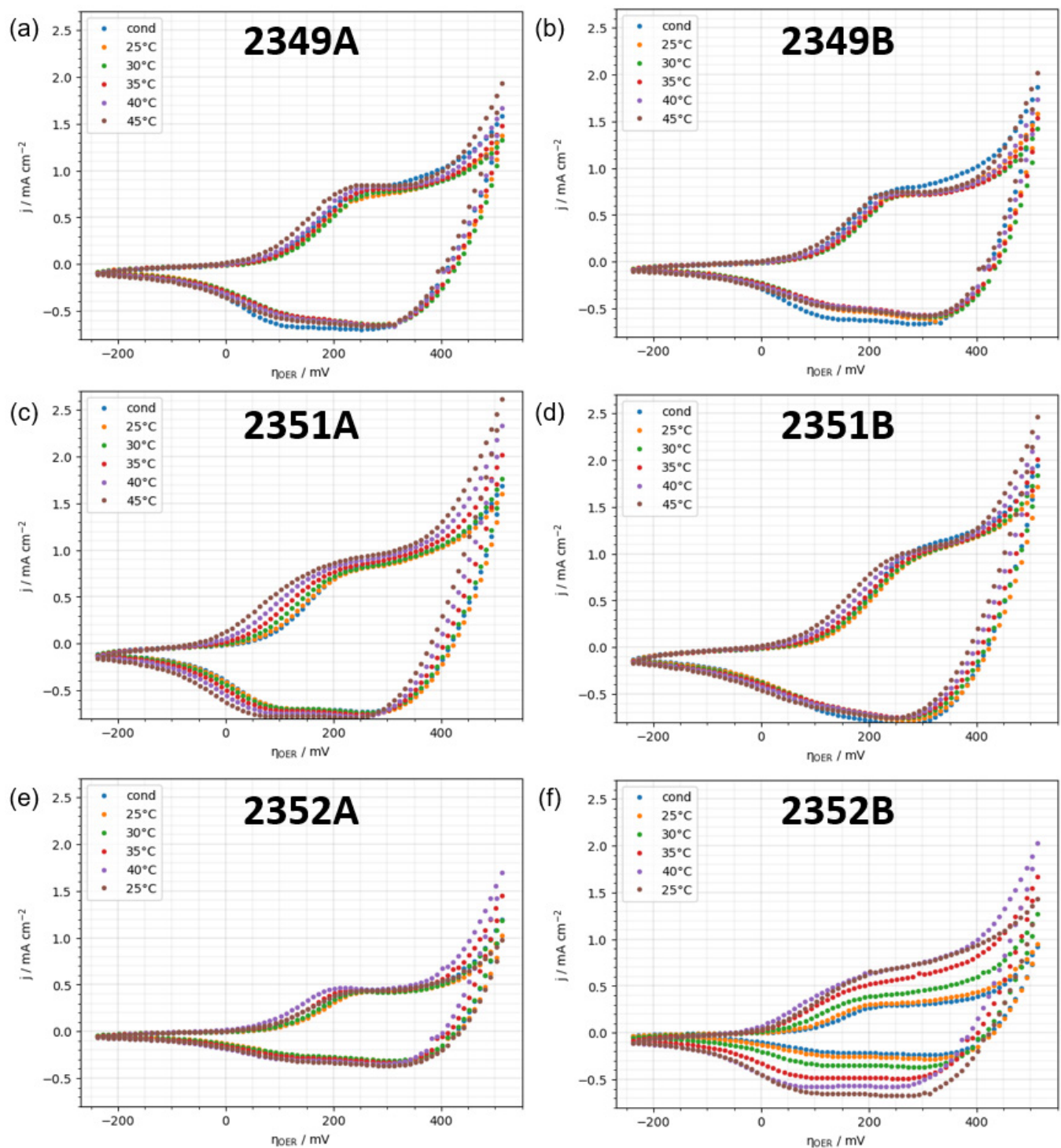

**Figure S6:** last cycle of each CV performed at the end of each iteration. Conditioning was also at 25°C.

## S7: Filtering - additional plots

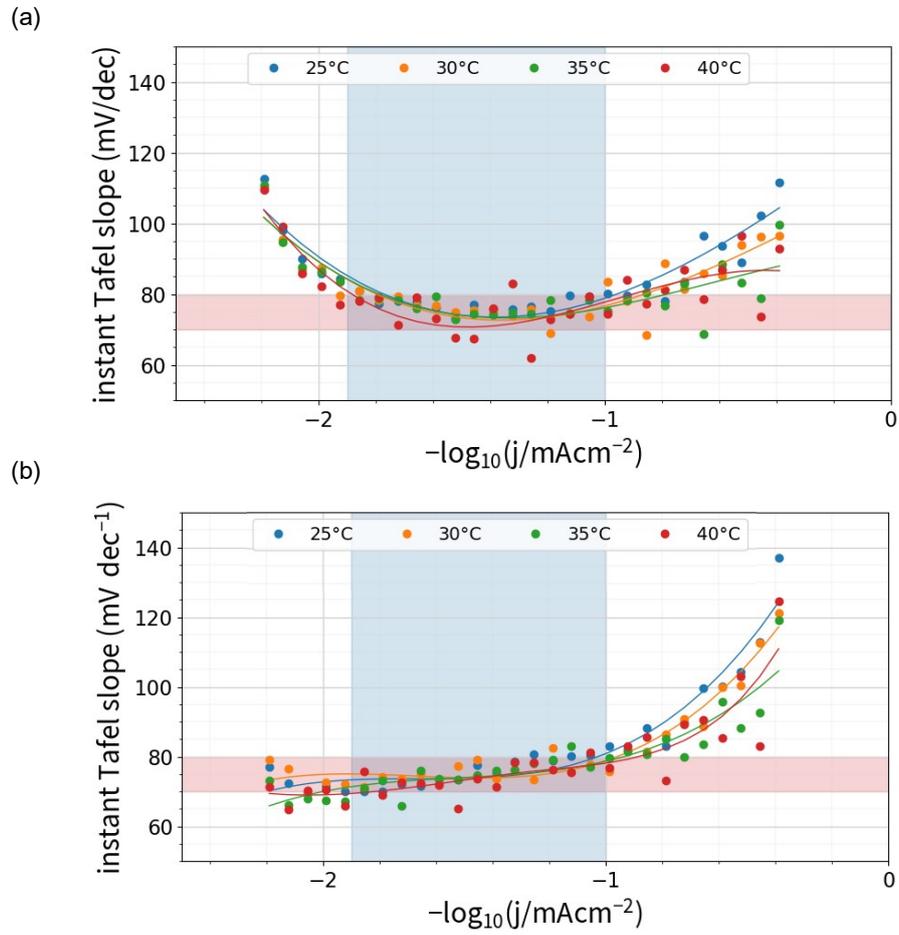

**Figure S7:** Instant Tafel slope value obtained as a discretization of the derivative of η with respect of $\log_{10}(j)$ at a constant temperature: $b = (\partial\eta/\partial \log_{10}(j))_T$ and cubic spline for each curve. This plot was used to determine which were the regions with Tafel behavior for: (a) SSP and (b) PEIS approaches (highlighted in blue and red). These plots show dataset EXP2349A.

# S8: Fitting results

## (A) SSP approach plots

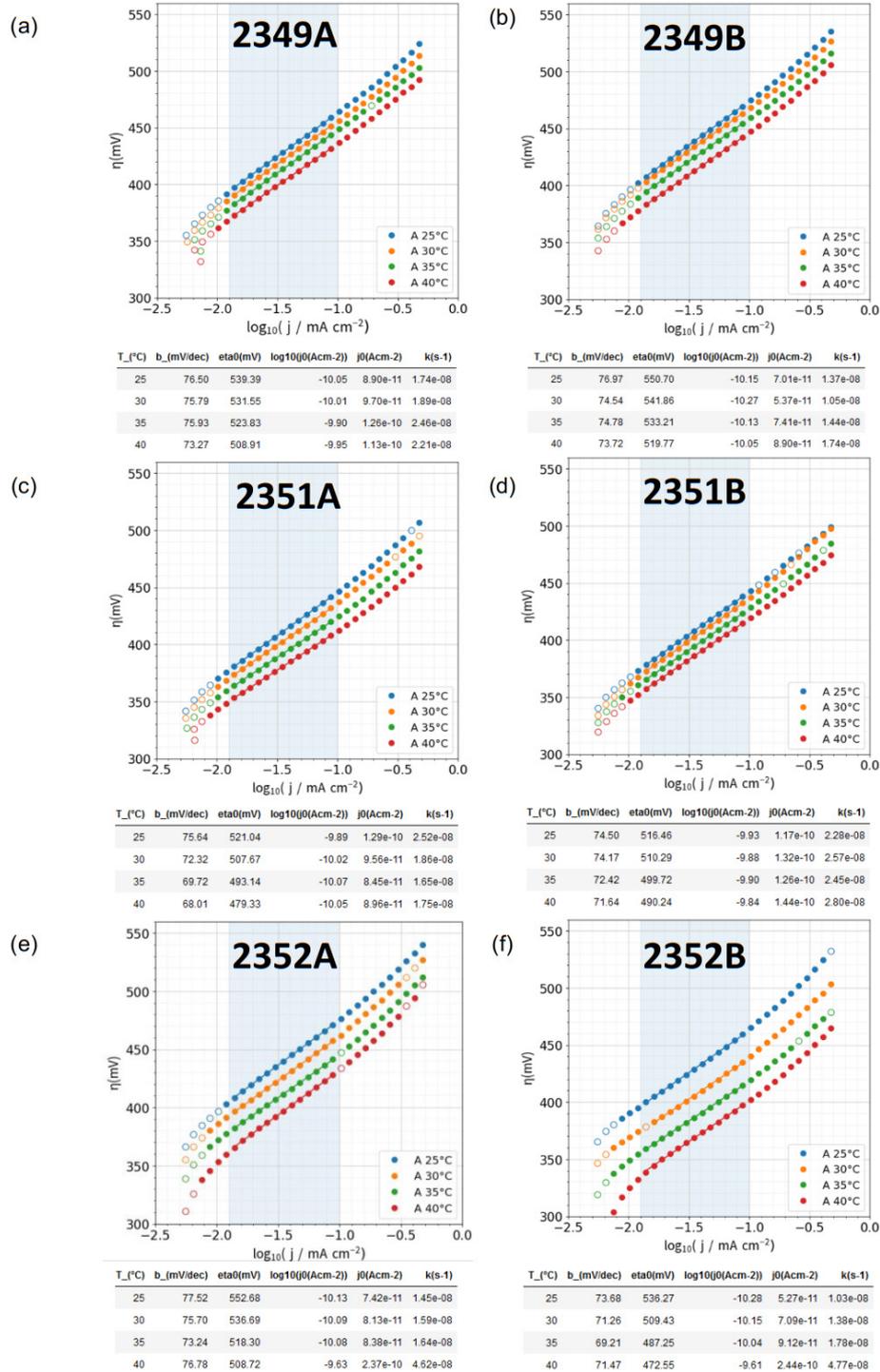

**Figure S8:** Tafel plots at different temperatures using SSP approach. Table columns are: temperature T (ºC), Tafel slope b (mV dec$^{-1}$), intersect with Y axis $\eta_0$ (mV), intersect with X axis $\log_{10}(j^0)$ (A cm$^{-2}$, logarithmic value and then decimal value), kinetic constant extrapolated to zero overpotential k$^0$ (s$^{-1}$). Nomenclature: ● fulfills Criterion 1 (potential drift < 1.10 mV). ○: does not fulfill neither Criteria 1. Points fulfilling Criterion 3 (instant Tafel slope of 75±5 mV dec$^{-1}$) have placed inside the blue-shaded area, which indicate that they have Tafel behavior and will be used for subsequent stages. Points outside the blue area are also excluded from the next stage of the analysis (construction of Eyring-Polanyi bilinear plots).

### (B) PEIS approach plots

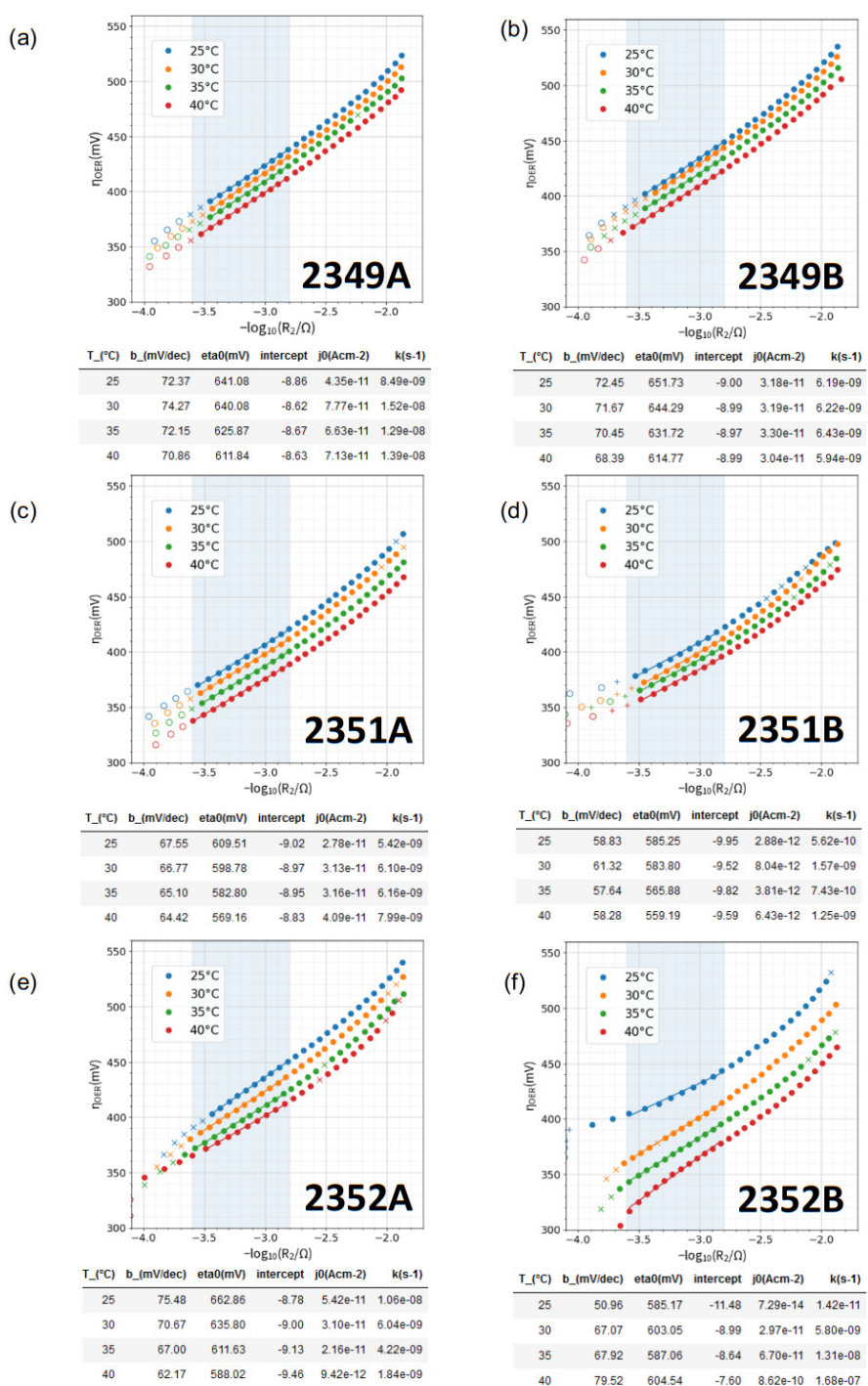

**Figure S9:** Tafel plots at different temperatures using PEIS approach. Table columns are: temperature T (°C), Tafel slope b (mV dec$^{-1}$), intersect with Y axis $\eta_0$ (mV), intersect with X axis -log$_{10}$(R$_2$) (with R2 in Ω cm$^{-2}$), exchange current density j$^0$ (A cm$^{-2}$), kinetic constant extrapolated to zero overpotential k$^0$ (s$^{-1}$). Nomenclature: ● fulfils both Criterion 1 and 2 (potential drift < 1.10 mV and fitting error smaller than 50% of the value, respectively). ○: does not fulfil neither Criteria 1 nor 2). ×: does not fulfil Criterion 1 but fulfils Criterion 2. +: fulfils Criterion 1 but does not fulfil Criterion 2. Points fulfilling Criterion 3 (instant Tafel slope of 75±5 mV dec$^{-1}$) have placed inside the blue-shaded area, which indicate that they have Tafel behavior and will be used for subsequent stages. Points outside the blue area are also excluded from the next stage of the analysis (construction of Eyring-Polanyi bilinear plots).

### S7: Fitting results

## (A) SSP Approach

**Table S4:** Results of independent fittings performed using SSP approach.

| dataset | beaker | α | $\Delta^{\ddagger}H°$(eV) | $\Delta^{\ddagger}S$(eVK$^{-1}$) | $T°\Delta^{\ddagger}S$(eV) | $\Delta^{\ddagger}G°$(eV) | $k^0$(s$^{-1}$) | $j^0$(Acm$^{-2}$) | A(V K$^{-1}$) | B(VK$^{-1}$) | C(VK$^{-1}$) | $C_1$(VK$^{-1}$) | $C_2$(VK$^{-1}$) |
|---|---|---|---|---|---|---|---|---|---|---|---|---|---|
| 2349 | A | 0.804 | 0.726 | -0.00165 | -0.492 | 1.22 | 1.61e-08 | 5.17e-11 | 0.000107 | 0.903 | 0.000121 | 0.00205 | -0.00193 |
| 2349 | B | 0.808 | 0.75 | -0.00161 | -0.479 | 1.23 | 1.06e-08 | 3.39e-11 | 0.000107 | 0.928 | 6.73e-05 | 0.00199 | -0.00192 |
| 2351 | A | 0.77 | 0.761 | -0.00144 | -0.429 | 1.19 | 4.69e-08 | 1.5e-10 | 0.000112 | 0.988 | -0.000147 | 0.00187 | -0.00202 |
| 2351 | B | 0.773 | 0.659 | -0.00178 | -0.53 | 1.19 | 4.92e-08 | 1.58e-10 | 0.000112 | 0.854 | 0.00029 | 0.0023 | -0.00201 |
| *2352** | *A* | *0.80* | *1.01* | *-0.00072* | *-0.22* | *1.22* | *1.22e-08* | *3.92e-08* | *0.000108* | *1.260* | *-0.00104* | *0.00091* | *-0.00194* |
| *2352** | *B* | *0.85* | *1.36* | *0.00041* | *0.12* | *1.24* | *7.58e-09* | *2.43e-08* | *0.000101* | *1.600* | *-0.00231* | *-0.00048* | *-0.00182* |

$C_1$ is the entropic and $C_2$ is the logarithmic term of C according to Equation 12h.
* dataset not used for statistical analysis.

**Table S5:** Statistics of the six independent fittings performed using SSP approach.

| | α | $\Delta^{\ddagger}H°$(eV) | $\Delta^{\ddagger}S$(eVK$^{-1}$) | $T°\Delta^{\ddagger}S$(eV) | $\Delta^{\ddagger}G°$(eV) | $k^0$(s$^{-1}$) | $j^0$(Acm$^{-2}$) | A(V K$^{-1}$) | B(VK$^{-1}$) | C(VK$^{-1}$) | $C_1$(VK$^{-1}$) | $C_2$(VK$^{-1}$) |
|---|---|---|---|---|---|---|---|---|---|---|---|---|
| count | 4 | 4 | 4 | 4 | 4 | 4 | 4 | 4 | 4 | 4 | 4 | 4 |
| mean | 0.789 | 0.724 | -0.00162 | -0.482 | 1.21 | 3.07e-08 | 9.84e-11 | 0.000109 | 0.918 | 8.3e-05 | 0.00205 | -0.00197 |
| std | 0.0201 | 0.0455 | 0.000139 | 0.0415 | 0.0198 | 2.02e-08 | 6.46e-11 | 2.78e-06 | 0.0559 | 0.00018 | 0.000182 | 5.01e-05 |
| min | 0.77 | 0.659 | -0.00178 | -0.53 | 1.19 | 1.06e-08 | 3.39e-11 | 0.000107 | 0.854 | -0.000147 | 0.00187 | -0.00202 |
| 25% | 0.772 | 0.709 | -0.00168 | -0.501 | 1.19 | 1.48e-08 | 4.73e-11 | 0.000107 | 0.89 | 1.38e-05 | 0.00196 | -0.00201 |
| 50% | 0.788 | 0.738 | -0.00163 | -0.485 | 1.2 | 3.15e-08 | 1.01e-10 | 0.000109 | 0.915 | 9.4e-05 | 0.00202 | -0.00197 |
| 75% | 0.805 | 0.752 | -0.00157 | -0.467 | 1.22 | 4.75e-08 | 1.52e-10 | 0.000112 | 0.943 | 0.000163 | 0.00211 | -0.00193 |
| max | 0.808 | 0.761 | -0.00144 | -0.429 | 1.23 | 4.92e-08 | 1.58e-10 | 0.000112 | 0.988 | 0.00029 | 0.0023 | -0.00192 |

## (B) PEIS Approach

**Table S6:** Results of the six independent fittings performed using PEIS approach.

| experiment | beaker | α | $\Delta^{\ddagger}H°$(eV) | $\Delta^{\ddagger}S$(eVK$^{-1}$) | $T°\Delta^{\ddagger}S$(eV) | $\Delta^{\ddagger}G°$(eV) | $k^0$(s$^{-1}$) | $j^0$(Acm$^{-2}$) | A'(K) | B'(KV$^{-1}$) | C'(-) | $C_1$' (-) | $C_2$'(-) |
|---|---|---|---|---|---|---|---|---|---|---|---|---|---|
| 2349 | A | 0.834 | 0.749 | -0.00149 | -0.443 | 1.19 | 4.5e-08 | 3.17e-11 | 9.68e+03 | -8.69e+03 | 8.34 | -17.2 | 25.6 |
| 2349 | B | 0.85 | 0.77 | -0.00147 | -0.438 | 1.21 | 2.4e-08 | 1.69e-11 | 9.86e+03 | -8.93e+03 | 8.55 | -17 | 25.6 |
| 2351 | A | 0.793 | 0.83 | -0.00111 | -0.332 | 1.16 | 1.42e-07 | 1e-10 | 9.2e+03 | -9.63e+03 | 12.6 | -12.9 | 25.5 |
| 2351 | B | 0.928 | 0.871 | -0.00119 | -0.356 | 1.23 | 1.14e-08 | 8.03e-12 | 1.08e+04 | -1.01e+04 | 11.8 | -13.9 | 25.7 |
| *2352** | *A* | *0.93* | *0.97* | *-0.00090* | *-0.27* | *1.24* | *6.38e-09* | *4.50e-09* | *1.08e+04* | *-1.13e+04* | *15.27* | *-10.41* | *25.68* |
| *2352** | *B* | *0.86* | *1.62* | *0.00137* | *0.41* | *1.21* | *2.03e-08* | *1.43e-08* | *1.00e+04* | *-1.88e+04* | *41.47* | *15.86* | *25.61* |

$C_1$' is the entropic and $C_2$' is the logarithmic term of C according to Equation 14h.
* dataset not used for statistical analysis.

**Table S7:** Statistics of independent fittings performed using PEIS approach.

| | α | $\Delta^{\ddagger}H°$(eV) | $\Delta^{\ddagger}S$(eVK$^{-1}$) | $T°\Delta^{\ddagger}S$(eV) | $\Delta^{\ddagger}G°$(eV) | $k^0$(s$^{-1}$) | $j^0$(Acm$^{-2}$) | A'(K) | B'(KV$^{-1}$) | C'(-) | $C_1$' (-) | $C_2$'(-) |
|---|---|---|---|---|---|---|---|---|---|---|---|---|
| count | 4 | 4 | 4 | 4 | 4 | 4 | 4 | 4 | 4 | 4 | 4 | 4 |
| mean | 0.851 | 0.805 | -0.00132 | -0.392 | 1.2 | 5.56e-08 | 3.92e-11 | 9.88e+03 | -9.34e+03 | 10.3 | -15.3 | 25.6 |
| std | 0.0565 | 0.0559 | 0.00019 | 0.0565 | 0.0274 | 5.93e-08 | 4.18e-11 | 655 | 649 | 2.2 | 2.2 | 0.0655 |
| min | 0.793 | 0.749 | -0.00149 | -0.443 | 1.16 | 1.14e-08 | 8.03e-12 | 9.2e+03 | -1.01e+04 | 8.34 | -17.2 | 25.5 |
| 25% | 0.824 | 0.764 | -0.00147 | -0.439 | 1.18 | 2.08e-08 | 1.47e-11 | 9.56e+03 | -9.75e+03 | 8.49 | -17.1 | 25.6 |
| 50% | 0.842 | 0.8 | -0.00133 | -0.397 | 1.2 | 3.45e-08 | 2.43e-11 | 9.77e+03 | -9.28e+03 | 10.2 | -15.4 | 25.6 |
| 75% | 0.869 | 0.84 | -0.00117 | -0.35 | 1.21 | 6.93e-08 | 4.88e-11 | 1.01e+04 | -8.87e+03 | 12 | -13.6 | 25.6 |
| max | 0.928 | 0.871 | -0.00111 | -0.332 | 1.23 | 1.42e-07 | 1e-10 | 1.08e+04 | -8.69e+03 | 12.6 | -12.9 | 25.7 |

## S8: Symbols meaning

**Table S8.** Symbols used in this work and their meaning.
(Extended version, includes symbols used in SI)

| Symbol | Meaning | Value | Unit |
|---|---|---|---|
| $A$ | geometric area of the WE | | $cm^2$ |
| $a$ | natural logarithm of 10 | $ln(10)$ | $-$ |
| $a_{H^+}$ | activity of protons | | $-$ |
| $a_{H_2O}$ | activity of water | | $-$ |
| $a_4$ | Constant phase element exponential parameter (Scheme S2) | | $-$ |
| $b_i$ | Tafel slope related to TS$_i$ | | $mV\ dec^{-1}$ |
| $C_1$ | bilinear fitting constant for galvanostatic SSP (x slope) | | $V$ |
| $C_1'$ | bilinear fitting constant used for PEIS (x' slope) | | $K$ |
| $C_2$ | bilinear fitting constant for galvanostatic SSP (y slope) | | $V\ K^{-1}$ |
| $C_2'$ | bilinear fitting constant used for PEIS (y' slope) | | $K\ V^{-1}$ |
| $C_3$ | bilinear fitting constant for galvanostatic SSP (z intercept) | | $V\ K^{-1}$ |
| $C_3'$ | bilinear fitting constant used for PEIS (z' axis intercept) | | $-$ |
| $E$ | equilibrium potential | | $V\ vs\ RE$ |
| $E_{OER}°$ | OER equilibrium potential (standard activities, T and p) | | $V\ vs\ RE$ |
| $e$ | elementary charge unit | $1.60 \cdot 10^{-19}$ | $C$ |
| $F$ | faraday constant | $96\ 485$ | $C\ mol^{-1}$ |
| $G$ | Gibbs free energy | | $eV$ |
| $G(\tau)$ | Transfer function | | $\Omega\ s$ |
| $h$ | Planck constant | $4.14 \cdot 10^{-15}$ | $eV\ s$ |
| $i$ | current | | $\mu A$ |
| $i_{OER}$ | faradaic current related to OER | | $\mu A$ |
| $j$ | current density | | $\mu A\ cm^{-2}$ |
| $j_{OER}$ | faradaic current density related to OER | | $\mu A\ cm^{-2}$ |
| $j^0$ | exchange current density (extrapolated to η = 0) | | $\mu A\ cm^{-2}$ |
| $k$ | heterogeneous kinetic constant | | $s^{-1}$ |
| $k^0$ | heterogeneous kinetic constant (extrapolated to η = 0) | | $s^{-1}$ |
| $k_B$ | Boltzmann constant | $8.62 \cdot 10^{-5}$ | $eV\ K^{-1}$ |
| $p_{O_2}$ | Oxygen partial pressure | | $atm$ |
| $R$ | gas constant | $8.31$ | $J\ mol^{-1}K^{-1}$ |
| $R_1$ | uncompensated resistance from circuit fitting (Scheme S2) | | $\Omega$ |
| $R_2$ | polarisation resistance from circuit fitting (Scheme S2) | | $\Omega$ |
| $R_p$ | polarization resistance | | $\Omega$ |
| $\tilde{R}_p$ | polarization resistance (normalized by area unit) | | $\Omega\ cm^{-2}$ |
| $R_u$ | uncompensated resistance | | $\Omega$ |
| $r$ | heterogeneous reaction rate | | $mol\ cm^{-2}s^{-1}$ |
| $T$ | temperature | | $K$ |
| $x$ | auxiliary variable for galvanostatic PEIS fitting | | $K^{-1}$ |
| $x'$ | auxiliary variable for galvanostatic SSP fitting | | $K^{-1}$ |
| $Y_3$ | Constant phase element parameter (Scheme S2) | | $s^{a_4}\ \Omega^{-1}$ |
| $y$ | auxiliary variable for galvanostatic PEIS fitting | | - |
| $y'$ | auxiliary variable for galvanostatic SSP fitting | | - |
| $z$ | auxiliary variable for galvanostatic PEIS fitting | | $V\ K^{-1}$ |
| $z'$ | auxiliary variable for galvanostatic SSP fitting | | $V\ K^{-1}$ |
| $z_e$ | number of electrons transferred per OER turnover | 4 | |
| $\Gamma_{act}$ | surface density of active sites | | $cm^{-2}$ |
| $\Delta^{\ddagger}G^0$ | Gibbs free energy of the RDS at η = 0 | | $eV$ |
| $\Delta^{\ddagger}G$ | Gibbs free energy of the RDS at a given η | | $eV$ |
| $\Delta^{\ddagger}H^0$ | enthalpy of the RDS at zero overpotential | | $eV$ |
| $\Delta^{\ddagger}H^0$ | enthalpy of the RDS at a given η | | $eV$ |
| $\Delta^{\ddagger}S$ | entropy of the RDS (invariant with η) | | $eV\ K^{-1}$ |
| $\alpha_{RDS}$ | transfer coefficient of the RDS | | $-$ |
| $\alpha_i$ | transfer coefficient of the i-th electron transfer | | $-$ |
| $\phi$ | WE potential after iRu correction | | $V\ vs\ RE$ |
| $\eta$ | overpotential after iRu correction | | $mV$ |
| $\eta'$ | overpotential without iRu correction | | $mV$ |
| $\eta_{theo}$ | theoretical overpotential | | $mV$ |
| $\kappa$ | transmission coefficient (TST) | | $-$ |
| $\tau$ | time constant | | $s$ |
| $\upsilon$ | scan rate | | $mV\ s^{-1}$ |
| $\chi^2$ | goodness of fit | | $-$ |